\documentclass[useAMS,usenatbib,A4]{mn2e}
\usepackage{psfig} 
\usepackage[dvips]{graphicx}
\usepackage{natbib}
\usepackage{epsf}

\def\plotone#1{\centering \leavevmode
\epsfxsize=\columnwidth \epsfbox{#1}}

\def\mnras{MNRAS}
\def\apj{ApJ}
\def\aap{A\&A}

\def\nat{Nature}

\title[Power spectrum of data with gaps]{A Mexican Hat with holes: calculating low
  resolution power spectra from data with gaps} 
\author[P. Ar\'evalo et al.]{P. Ar\'evalo$^{1,2}$\thanks{E-mail: parevalo@unab.cl}, E. Churazov$^{2,3}$, I. Zhuravleva$^{2}$, C. Hern\'andez-Monteagudo$^{2,4}$,\and M. Revnivtsev$^{3}$\\ 
$^1$Departamento de Ciencias Fisicas, Universidad Andres Bello, Av. Republica 252, Santiago, Chile\\
$^2$Max-Planck-Institut f\"ur Astrophysik, Karl-Schwarzschild-Strasse 1, 85741 Garching, Germany\\
$^3$Space Research Institute (IKI), Russian Academy of Sciences, Profsoyuznaya 84/32, 117997 Moscow, Russia\\
$^4$Centro de Estudios de F\'isica del Cosmos de Arag\'on (CEFCA), Plaza San Juan, 1, Planta 2, E-44001 Teruel, Spain
}

\begin{document}
\date{Received /Accepted}
\pagerange{\pageref{firstpage}--\pageref{lastpage}} \pubyear{2009}

\maketitle
\label{firstpage}

 \begin{abstract}
A simple method for calculating a low-resolution power spectrum from
data with gaps is described. The method is a modification of the
$\Delta$-variance method previously described by Stutzki and
Ossenkopf. A Mexican Hat filter is used to single out fluctuations at
a given spatial scale and the variance of the convolved image is
calculated. The gaps in the image, defined by the mask, are corrected
for by representing the Mexican Hat filter as a difference between two
Gaussian filters with slightly different widths, convolving the image
and mask with these filters and dividing the results before
calculating the final filtered image. This method cleanly compensates
for data gaps even if these have complicated shapes and cover a
significant fraction of the data. The method was developed to deal
with problematic 2D images, where irregular detector edges and masking
of contaminating sources compromise the power spectrum estimates, but
it can also be straightforwardly applied to 1D timing analysis or 3D
data cubes from numerical simulations.
\end{abstract}

\section{Introduction}

\label{sec:introduction}

The calculation of the Power spectrum through direct Fourier transform
of 2D data in astrophysics is often hampered by two problems.

\begin{description}
\item[{ I. }] Images may have irregular boundaries or parts of the
  image are missing. In many cases masks are applied to remove
  contaminating foreground sources, creating holes in the data that
  are hard to correct for in Fourier analysis. Such problems arise,
  for example, when angular fluctuations of a diffuse emission are
  analyzed and a number of compact sources have to be excised from the
  image \citep[e.g.][]{2012MNRAS.tmp.2290C}.

\item[{ II. }] Another problem often encountered when dealing with
  limited data sets is the presence of large scale structures, which
  are not fully covered by the image. The large scale power can leak
  into the observable Fourier frequency range, distorting the measured
  spectrum. This often occurs in 1D timing analysis when the noise
  process monitored has power on timescales longer than the total
  length of the time series \citep[e.g.][]{rednoise}, or in 3D data
  cubes of hydrodynamical simulations for example, when characterizing
  the turbulent velocity field in a sub-volume of a larger simulated
  volume \citep[e.g.][]{hydrosim}.
\end{description}

\begin{figure*}
\includegraphics[width=0.48\textwidth]{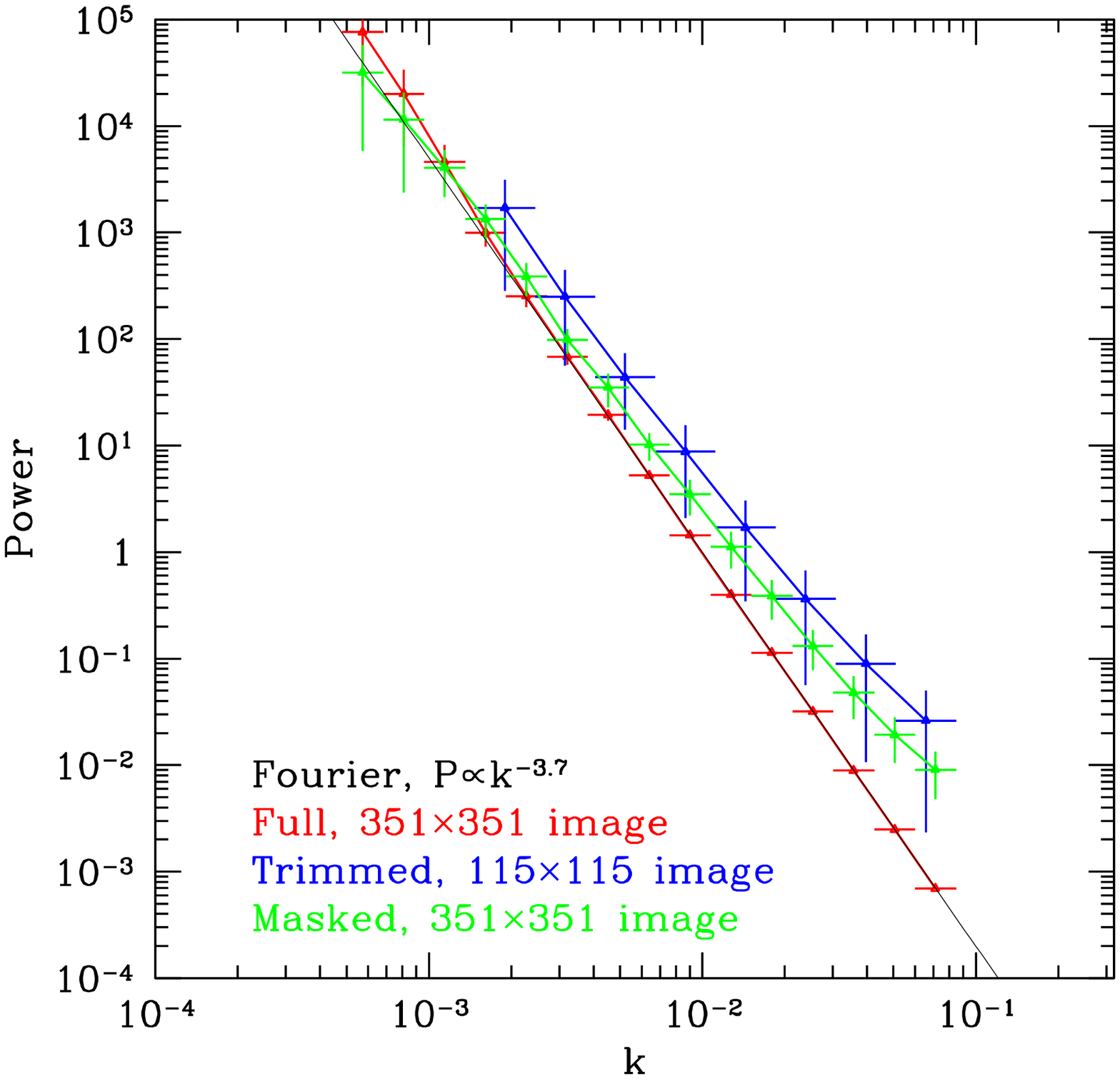}
\includegraphics[width=0.48\textwidth]{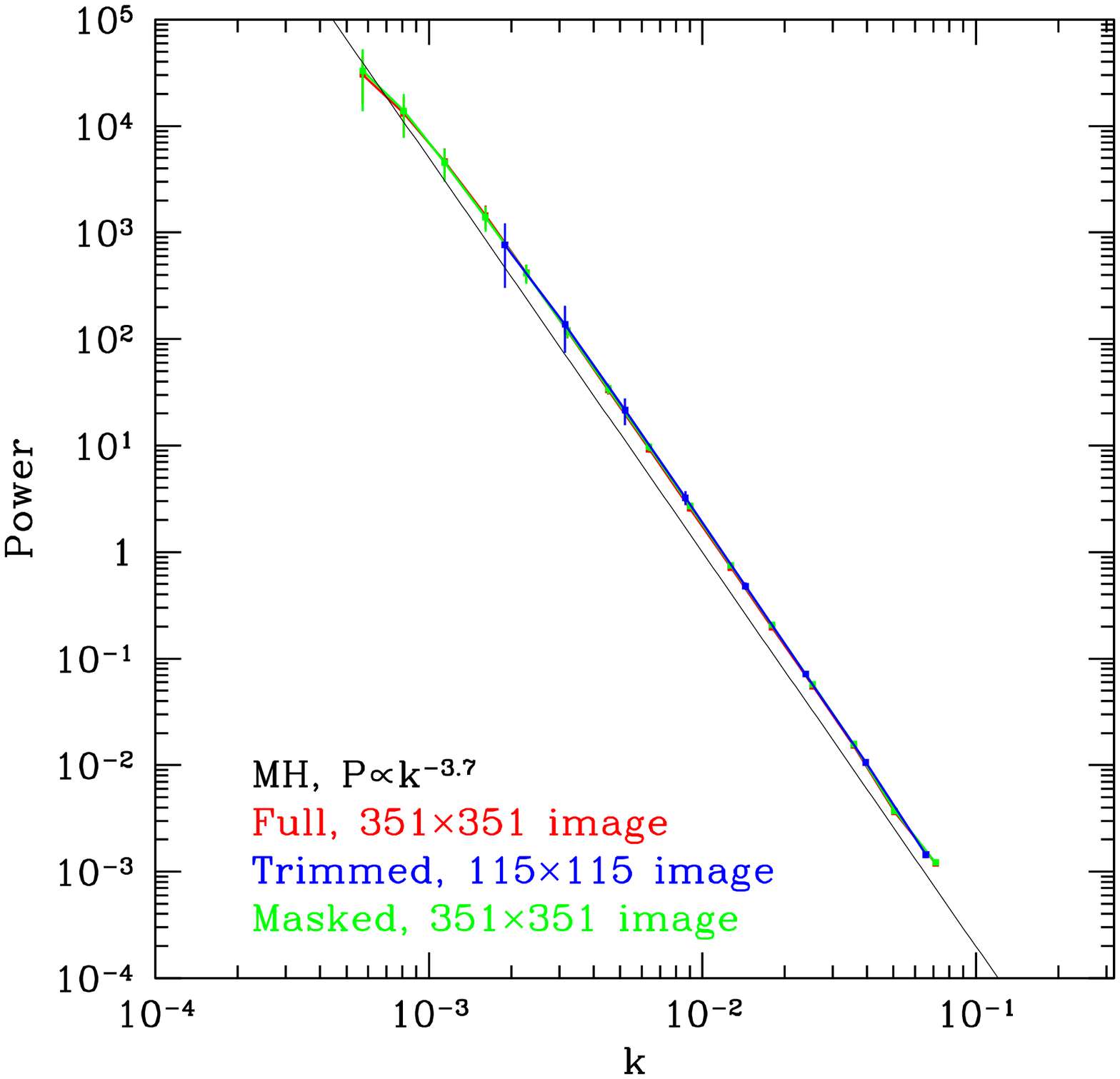}
\caption{{ Left:} Fourier power density spectrum (PDS) calculated for
  a 2D image. As input a $351\times 351$ pixel image with steep PDS
  ($\propto k^{-11/3}$) was used (shown by the black line). The
  spectra were obtained by averaging the power spectra, recovered from
  50 random realizations of images with the same power spectrum. The
  errorbars were estimated from the scatter between recovered PDS.
  The red points show the Fourier PDS for the whole image, blue points
  show the PDS calculated for a small section ($115 \times 115$ pixel)
  of the original image and green points show the case when about 25\%
  of the data in the original $351\times 351$ pixel image are missing
  and replaced with zero. Clearly both the normalization and the shape
  of the PDS are modified when parts of the data are missing or
  significant power is present on spatial scales larger than the size
  of the image (case of image subsection).  { Right:} Power estimated
  using the Mexican Hat filter, corrected for the gaps in the
  data. The same set of data is used as in the left panel.  Clearly
  the result is insensitive to the presence of the data gaps. There is
  a weak bias in the normalization, which is discussed in the text and
  in the Appendix \ref{ap:bias}.}
\label{fig:intro}
\end{figure*}

These problems are illustrated in the left panel of
Fig.\ref{fig:intro}, which shows the Fourier power density spectrum
(PDS) calculated for a 2D image. As input an image with steep PDS
$\propto k^{-11/3}$ was used. The red points show the Fourier PDS
for the whole image, blue points show the PDS calculated for a section
one third of the linear size of the original image and green points show the
case when about 25\% of the data in the original image are missing and
replaced with zero. Changes in the slope and normalization are readily
visible for blue and green curves.

A practical way to deal with these problems has been suggested in a
series of papers by \citet{stutzki98,bensch01,ossenkopf08}. Their
 $\Delta$-variance method preferentially selects fluctuations at
a given spatial scale $\sigma$ by convolving an image with two filters
- a compact ``core'' filter and a more extended ``annulus''. The core
filter has a characteristic size $\sim\sigma$ and both filters are
normalized to unity. The difference between the images convolved with
these two filters is an image where all fluctuations with
sizes much larger or much smaller than $\sigma$ are
suppressed. Therefore the variance of the resulting image provides a
measure of a typical amplitude of fluctuations of size $\sim\sigma$ in
the original image. Introduction of a mask helps to deal with the
boundaries or data gaps.

In this paper we discuss a simple method, based on the
$\Delta$-variance approach to compute a
low resolution power spectrum (strictly speaking the convolution of the
true power spectrum with a broad filter) even when a large fraction of
the original data is missing. The right panel in Fig.~\ref{fig:intro}
shows power spectra computed with our method for the same images used
in the left panel. Compared to the original
$\Delta$-variance method, our approach uses a different implementation
of the filter, which simplifies the procedure, while cleanly
compensating for missing data. We further test this particular implementation
for a number of potential applications -- e.g. measuring the power
spectrum of the surface brightness
fluctuations of the X-ray images of galaxy clusters, or characterizing
the power spectrum of a turbulent velocity field in simulations.

Many other methods have been devised to deal with data with
gaps. Notably, multi-taper analysis techniques have been developed to
estimate the power spectra of one-dimensional \citep{MTthomson} and
multidimensional \citep{hanssen} data in Cartesian coordinates and on
the sphere \citep{wieczorek}. The choice of taper functions is adapted
to the type of data to be treated and the number of tapers is chosen
to balance the bias and variance in the resulting power
spectrum. These methods can recover very high resolution power
spectra, suppressing the power leakage produced by the finite data
length and are ideally suited to well sampled data where only few or
no points are missing. Severe gaps in the data complicate the
application of these methods however.  \citet{fodor98} deal with large
gaps by calculating separate power spectra for each segment and
averaging together the results, which is not easily applicable to data
sets with varying data and gap lengths or data in more than 1D.
\citet{fodor} uses multi-tapering methods to compute the power
spectrum of complete data sets with few  small gaps, which requires
the calculation of optimized taper functions for the precise structure
of gaps in the data. Although this method does a very good job at
recovering the power spectral shape, it requires long and complex
calculations which are not directly transferable between data sets
with different sampling patterns and was only tested for cases where
gaps cover a small fraction (e.g. 5\%) of the data.  

The method we discuss in this paper is simple, robust and
computationally fast. It has no tuning parameters and the
interpretation of the resulting power spectrum is
straightforward. Most importantly, it can be applied without
modifications to the data severely affected by gaps of different
sizes.   The trade-off is its low spectral resolution, so it is 
useful for cases when the underlying power spectrum is a smooth
function of a wavenumber/frequency.  Possible applications in
astrophysics include  aperiodic variability patterns normally found
in AGN light curves; analysis of fluctuations in 2D images, e.g. maps
of molecular lines, X-ray images, Faraday Rotation Measure maps
defined in irregularly shaped regions; characterization of 3D
density or velocity fields in numerical simulations.  All  these
cases are often affected to a varying degree by gaps in the data.
These gaps arise  mainly due to time constraints in the
observations in time series, co-adding of several 2D images with
different orientations  and excision of contaminating sources and
limited computational volumes in 3D simulations. 

We demonstrate that for data sets of sufficiently large dynamic range
the proposed method recovers well the overall shape and
normalization of the spectrum  even when gaps occupy large
fraction of the data set.

In terms of uncertainties in the power spectrum estimation at a
given frequency, the method is analogous to the regular Fourier power
spectrum, for data sets without gaps, provided that similar binning of
the Fourier powers is made in the frequency space.

In the following section we will describe the method
(Sec.~\ref{sec:method}) and demonstrate its ability to recover the
original power spectrum from simulated 2D images with gaps in
Sec.~\ref{tests}.   We also apply the method in spherical
coordinates  and use it to evaluate the power spectrum for a
characteristic case of CMB data analysis in Sec. \ref{sec:CMB}. We
explore the properties and applications of the method to 3D data cubes
in Sec.~\ref{sec:3D} and 1D time series in Sec.~\ref{sec:timing}. 
  Expected scatter and errors are discussed in Sec.~\ref{sec:error}
  and we summarize our conclusions in Sec.~\ref{sec:conclusion}.

\section{Method}
\label{sec:method}
We consider an isotropic homogeneous random field, so that it can be
characterized by a power spectrum $P(k)$, where $k$ is a
scalar. Instead of calculating the ``true'' power spectrum $P(k)$,
where $k$ is the wave number, we want to evaluate the amplitude of
fluctuations for a broad interval of wave numbers $\Delta k\sim k$. We
allow for gaps in the data and want to recover  the correct shape and normalization of the power spectrum,
provided that $P(k)$ does not contain sharp features.
 
The method consists of few simple steps for any given spatial scale
$\sigma$:
\begin{description}
\item[(i)] the original image is convolved with two filters
(e.g. Gaussian) having different smoothing lengths
  $\sigma_1=\sigma/\sqrt{1+\epsilon}$ and
  $\sigma_2=\sigma\sqrt{1+\epsilon}$, where $\epsilon \ll 1$.
\item[(ii)] convolved images are corrected for the data gaps (see
\S\ref{sec:method_wg}).
\item[(iii)] the difference of two images is calculated. This
  difference image is dominated by fluctuations at scales $\sim \sigma$.
\item[(iv)] The variance of the resulting image is calculated and
  re-casted into an estimate of the power. 
\end{description}
The variance values are collected as a function of length scale or,
correspondingly, wave number $k_r\propto 1/\sigma$ to produce the
power spectrum $\tilde{P}(k_r)$. The tilde is added to distinguish
the result from the true power spectrum.

We first consider the case of data without gaps
(\S\ref{sec:method_ng}, steps i,iii and iv) and then discuss the
procedure of correcting for gaps (\S\ref{sec:method_wg}) using a Mexican
Hat filter as an example.

\subsection{Data without gaps}
\label{sec:method_ng}
\begin{figure}
\includegraphics[width=0.80\columnwidth,angle=270]{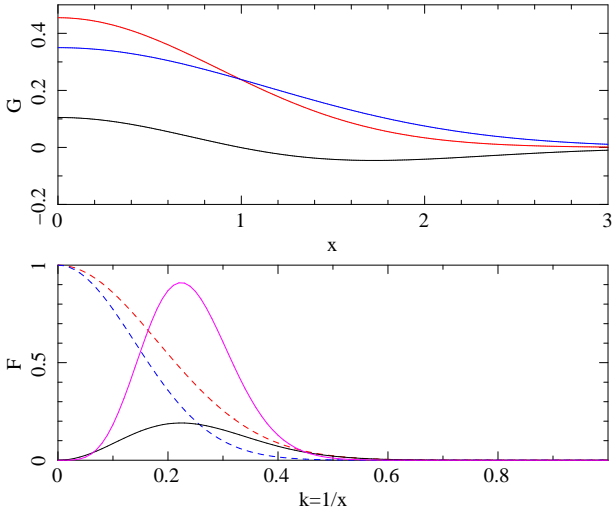}
\caption{{\em Top panel:} 2 Gaussians of $\sigma \sim 1$ used for
  filtering (red, blue) and their difference (black). {\em Bottom
  panel:} Fourier transforms of the Gaussian functions shown above
  (red, blue), their difference (black) and the difference squared,
  amplified for clarity (magenta). The magenta curve is the filter
  effectively applied on the power spectrum of the image treated,
  therefore the variance of the filtered image is normalized by the
  area under this curve. The transmitted power peaks at $k=0.225/\sigma$.}
\label{fig:filter}
\end{figure}

To isolate structures of a characteristic length-scale, we first smooth
the image $I$ with two Gaussian filters (for simplicity we consider 1D case):
\begin{equation}
G_{\sigma}(x)=\frac{1}{(2\pi\sigma^2)^{1/2}}e^{-\frac{x^2}{2\sigma^2}}
\end{equation}
 of slightly different widths $\sigma_1=\sigma/\sqrt{1+\epsilon}$ and
 $\sigma_2=\sigma \sqrt{1+\epsilon}$, where $\epsilon \ll 1$. The top
 panel in Fig. \ref{fig:filter} shows an example of two such Gaussian
 filters, normalized to unity, together with their difference. For
 clarity only, in the Figure we use $\epsilon\sim 0.25$. After
 convolving the image with each of these filters, both resulting
 images $I_1$ and $I_2$ will retain structures larger than $\sigma$
 and lose structures smaller than $\sigma$ so the difference image
 $\displaystyle I_1-I_2$ will predominantly contain structures with
 the characteristic scales $\sim\sigma$.

For $\epsilon \rightarrow 0$ the resulting filter 
\begin{eqnarray}
F(x)=G_{\sigma_1}(x)-G_{\sigma_2}(x)\propto \frac{\partial
  G_{\sigma}(x)}{\partial\sigma}(\sigma_2-\sigma_1)\propto \nonumber \\
 \propto \epsilon\left [ 1 -\frac{x^2}{\sigma^2}\right] e^{-\frac{x^2}{2\sigma^2}},
\label{eq:fs0}
\end{eqnarray}
which is the familiar Mexican Hat filter. Obviously the shape of the
filter does not depend on $\epsilon$ in the limit of $\epsilon
\rightarrow 0$. In practice we use $\epsilon=10^{-3}$.
 
In terms of power spectra, the above procedure is equivalent to
multiplying the original power spectrum of the data by the power
spectrum of the filter, shown in the bottom panel of
Fig. \ref{fig:filter}. This panel shows the Fourier transforms of
the two Gaussian filters, which are of course also Gaussian, and 
their difference. The square of the difference, is  the filter
 power spectrum. As shown in Appendix \ref{ap:power},  the peak
of the filter power is at $k_r=0.225/\sigma$ and its width is $\sim
1.155~k_r$, independent of $\epsilon$, as long as $\epsilon \ll
1$. Here and below we adopt the relation between the spatial scale
$x$ and a wavenumber $k$ in a form $k=1/x$ without a factor $2\pi$.
 The resulting filter is relatively broad and it cannot be made
narrower by using smaller values of $\epsilon$. This has the
implication that sharp features in the power spectrum will be
smeared out, so our aim is to recover the broad band shape and
normalization of the power spectrum but not narrow features. The
choice of the Gaussian filters is twofold. Firstly, they provide
a good balance in terms of the width in real and frequency space. The
width in the real space is important when gaps or edges are present in
the data (see next Section). A small width implies that the region
affected by a given gap does not extend over the whole
image. Secondly, the use of Gaussian filter is computationally
convenient, since n-dimensional convolution with a Gaussian in a real
space can be easily factorized into 1D convolutions in each
dimension.

\begin{figure*}
\includegraphics[width=2.0\columnwidth,angle=0]{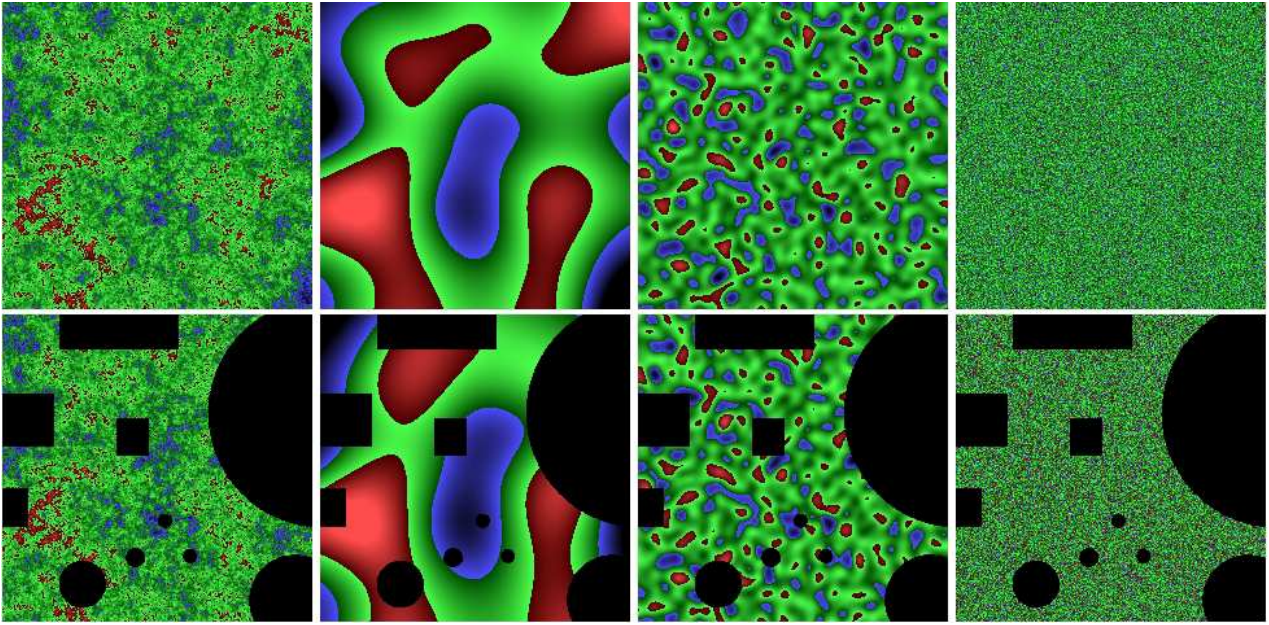}
\caption{{ {\em Top row:} simulated image with power law
  shaped PDS, each subsequent panel shows the result of filtering the
  image at a different spatial scale $k$ as described in the text,
  from left to right: $k^{-1}/L=0.6, 0.08$ and $0.02$,
  where $L$ is the size of the image. {\em Bottom row:}
  original image multiplied by the mask, i.e. $I\times M$. Each
  subsequent panel shows filtered images calculated from the same
  simulated masked image after application of eq.~\ref{eq:ic} and
  multiplication by the original mask, i.e. $I_c(k_r)\times M$. No
  spurious peaks or dips are produced around the masked regions or near
  image edges. These
  parts of the filtered images where $M=1$ are used to calculate the
  variance for a given $k_r$.}}
\label{mask}
\end{figure*}

Thus the difference between two convolved images, which we denote as
$I(k_r)$ is
\begin{equation}
I(k_r)=I_1-I_2 \equiv G_{\sigma_1}*I-G_{\sigma_2}*I\equiv F*I,
\end{equation}
can be used to calculate the variance at scales $\displaystyle \sim
1/k_r$ and evaluate $\tilde{P}(k_r)$ using a simple relation for the
normalization (see Appendix \ref{ap:power}).

\subsection{Data with gaps}
\label{sec:method_wg}
We now consider the image with gaps and introduce a mask $M$ such that
\begin{eqnarray}
M=\left\{\begin{array}{ll}
1 & {\rm where~} I {\rm ~is~defined} \\
0 & {\rm where~} I {\rm ~is~undefined} \\
\end{array}
\right.
\end{eqnarray}
Here ``undefined'' refers not only to gaps, but also to areas outside the image
boundaries. Essentially for an $n$-dimensional image we treat the whole
$n$-dimensional space outside the image boundaries as a data gap.
The image $I$ is also set to zero in the gaps and outside image
boundaries. Thus, one can write $I=M\times I_0$, where $I_0$ is the
true image without gaps, defined over the whole $n$-dimensional space.

Direct application of the filter $F$ described by eq.~\ref{eq:fs0} to
the image $I$ with gaps will produce many spurious structures, which
are difficult to correct for. However one can use the fact that the
$I(k_r)$ image can be represented as the difference of two smoothed
images. Consider, for instance, $\displaystyle
I_1=G_{\sigma_1}*I$. Convolving the image with gaps with a Gaussian
will still produce spurious features, but their amplitude can be
drastically reduced by dividing $I_1$ by the mask $M$,
convolved with the same Gaussian, to produce a corrected image
$\displaystyle I_{1,c}$:
\begin{equation}
I_{1,c}=\frac{I_1}{M_1}=\frac{G_{\sigma_1}*(M\times I_0)}{G_{\sigma_1}*M}.
\end{equation}
Note, that we make the convolution in the infinite $n$-dimensional
space without assumption of the image periodicity outside the image
boundaries. Intuitively the effect of the division by the convolved
mask is clear: the amplitude of
$I_1=G_{\sigma_1}*I=G_{\sigma_1}*(M\times I_0)$ is going to be lower close to
the gaps or close to image boundaries. The convolved mask $M_1=G_{\sigma_1}*M$
largely shares these properties and the ratio $\frac{I_1}{M_1}$ will
lack an obvious trend of the amplitude decrease near the gaps. 
The same argument applies to $\displaystyle
I_{2,c}=\frac{I_2}{M_2}$. Finally
\begin{equation}
I_c(k_r)=(I_{1,c}-I_{2,c})\times M=\left(\frac{G_{\sigma_1}*I}{G_{\sigma_1}*M}-\frac{G_{\sigma_2}*I}{G_{\sigma_2}*M}\right)\times M.
\label{eq:ic}
\end{equation}

The final step of the variance calculation is done only for the part
of the corrected image $I_c(k_r)$ where the mask $M=1$ (see Appendix
\ref{ap:power}).  In summary, the biases introduced by the shape of
the image boundaries and holes in the mask are corrected for by
subjecting the mask to the same smoothing procedure and then dividing
the smoothed image by the smoothed mask. Therefore, the flux lost by
smearing the image out of the mask boundaries is compensated by an
equal loss of mask area around the edges.  When approaching a gap,
  the convolved mask $G_{\sigma_1}*M$ decreases smoothly from a value
  of 1 before the gap to 0 well into the gap, provided that the gap
  is larger than the filter size. Therefore the convolved mask
  typically has a value of order 0.5 or larger at the edge of the
  original gap\footnote{ Unless we are dealing with an isolated data
    area, small compared to the size of the filter and surrounded by
    gaps.}. Since only unmasked parts of the image are used for the
  variance calculations, points where the denominators in
  Eq. \ref{eq:ic} vanish are automatically discarded.

Our approach is analogous to the $\Delta$-variance method of
\citet{stutzki98, bensch01} and \citet{ossenkopf08}.  In particular,
one of the suggested forms of the filters in \citet{ossenkopf08} (see
their equation 11) leads to a convolution of an image with a Mexican
Hat filter, in the limit of their parameter $v\rightarrow 1$. In
other words the difference between filters is the same as used
here. However the individual filters are different. \citet{ossenkopf08}
use a Gaussian as a ``core'' filter and a ring-like shape function
produced by a linear combination of two Gaussians, as a ``annulus'' filter. In our
implementation of the Mexican Hat filter the role of the core and
annulus filters (see \citealt{stutzki98,ossenkopf08} for definitions)
is played by two Gaussians with slightly different widths. Far from image gaps and boundaries
only the
difference between the core and annulus filters matters. But in the
presence of gaps or boundaries the functional forms of both filter
starts to be important.
Representing the Mexican Hat as a difference between two
Gaussians simplifies the whole procedure when correcting for the
missing data and ensures that the effect of gaps is almost identical
for $I_{1,c}$ and $I_{2,c}$. 

Figure \ref{mask} shows a simulated image with an isotropic power law
power spectrum of slope $\alpha=-2$. The top row in this figure shows
the image decomposition into components of different spatial
scales. The bottom row shows the same simulated image to which an
arbitrary mask is applied (left). Other images in the bottom row
show the decomposition performed on the \emph{masked}
image. Clearly, the method is quite insensitive to the presence of
the mask, as seen in Fig.~\ref{mask}. Indeed, the fluctuations on different
spatial scales can be recovered in the masked image without
introducing spurious structures near the edges or gaps.

\section{Test on simulated images}
\label{tests}
\begin{figure}
\begin{center}
\includegraphics[width=0.5\columnwidth,angle=270]{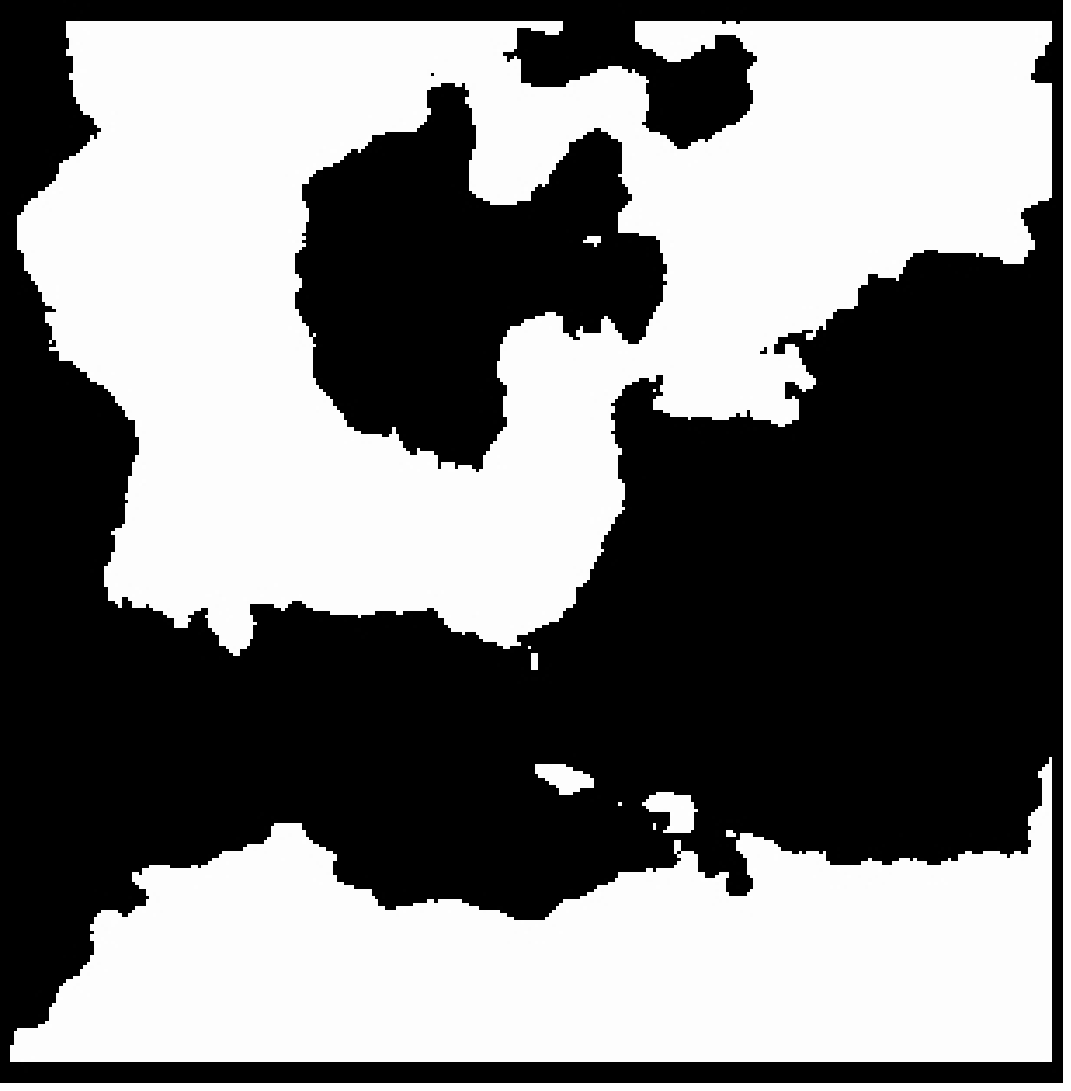}
\includegraphics[width=0.5\columnwidth,angle=270]{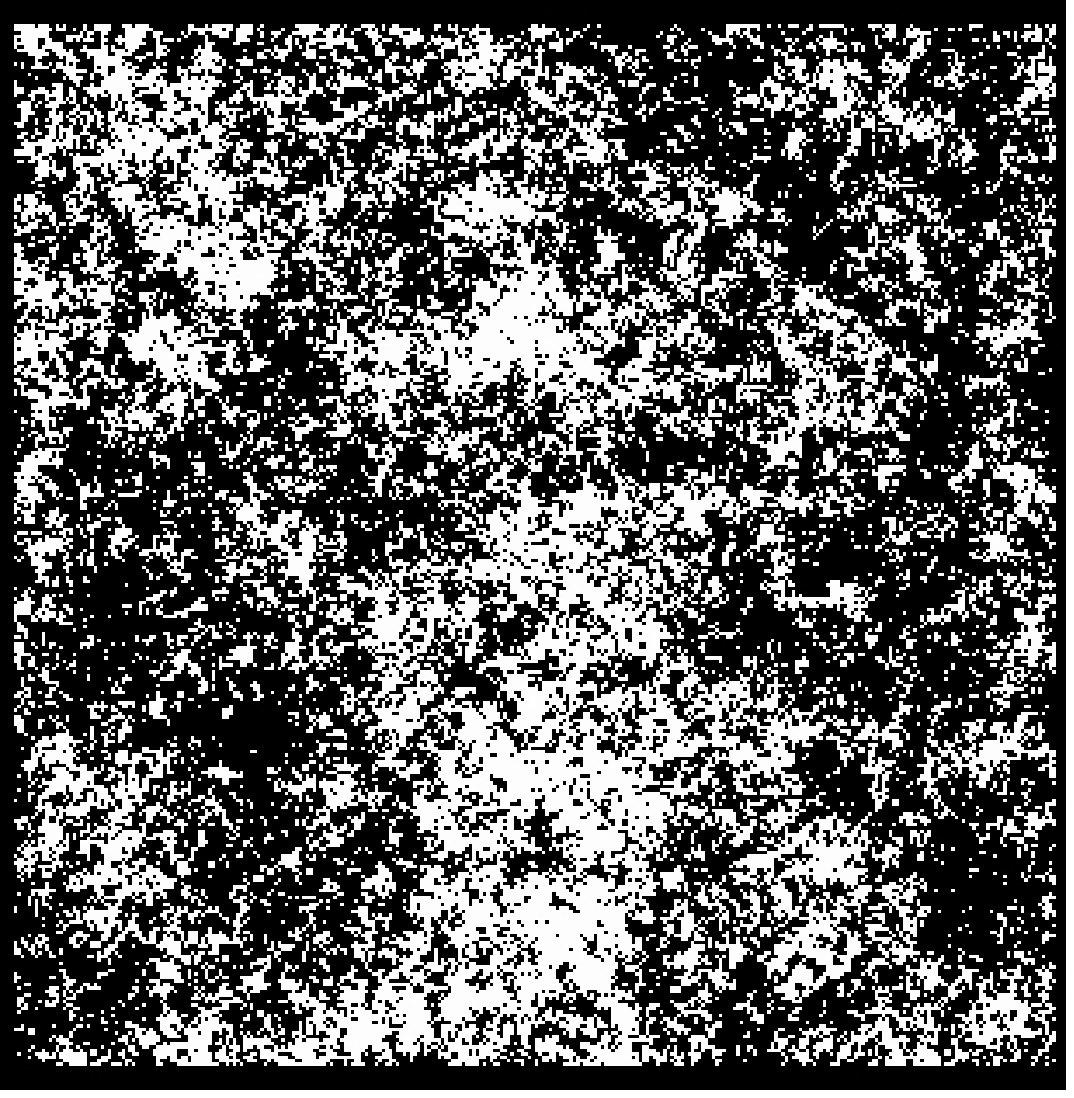}
\caption{Mask 1 (left) and mask 2 (right) applied to simulated 2D
  images before computing their $\tilde P(k_r)$ spectra. The black
  areas represent the discarded parts of the input images. Data gaps
  in real astronomical images  are usually less extreme, compared
  to mask 2.  We use this
  extreme example to test the  ability of the method to recover the input power
  spectrum even for images this heavily affected by gaps.}
\label{fig:2Dmask}
\end{center}
\end{figure}

To test the ability of the method to recover the original power
spectrum of an image, we generated a set of 2D Gaussian random
fields. We simulate the effect of irregular edge shapes and
excision of contaminating sources  by masking out regions
with different shapes and size scales.  The mask in Fig.~\ref{mask} is
typical for wide field images when different exposures are combined
and contaminating sources are excluded, creating irregular edges and
holes in the mosaiced image. For the tests below we use more complicated masks, shown in
Fig.~\ref{fig:2Dmask} to verify the performance of the method in more
extreme cases. The tests have also been run using the simple mask
shown in Fig.~\ref{mask}, recovering the input power spectra at least
as accurately.

{ We checked for overall biases in the
power spectrum shape and normalization by
generating images with randomized 
phases and amplitudes to simulate a Gaussian field
\citep[e.g.][]{timmer}. Images were constructed with power law power spectra of slopes
between 0 and --4 and, for each case, 100 
realisations were produced and masked by the patterns shown in
Fig.~\ref{fig:2Dmask}.}

\begin{figure}
\begin{center}
\includegraphics[width=1.5\columnwidth,bb=110 100 570 370,clip,angle=270]{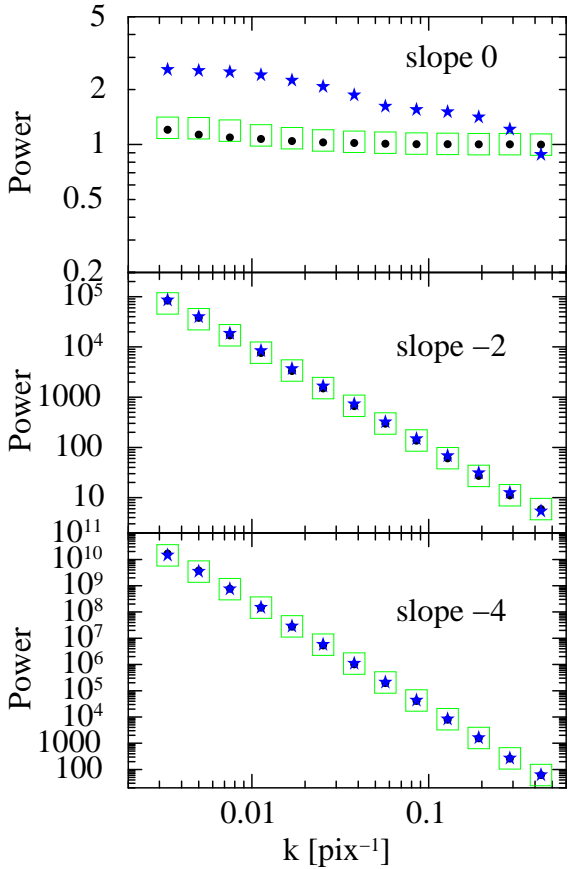}
\caption{ Recovered power spectrum of simulated images. The markers
  show averaged $\tilde P(k_r)$ spectrum over 100 realizations of
  simulated images. Black dots show the spectrum computed for unmasked
  images. Green squares and blue stars correspond to images masked
  using patters 1 and 2 (see Fig.~\ref{fig:2Dmask}).  The slopes of
  the input  power spectra are given in each panel. The input
  spectra are normalized to unity at $k=1$.}
\label{fig:mean_var}
\end{center}
\end{figure}

Figure \ref{fig:mean_var} shows the variance of the filtered images as
a function of their characteristic filter scale $k_r$, for 
different slopes of the power spectrum. Simulated images had a size of
$1024^2$ pixels and a central $300^2$ pixels section of each image was used
to calculate the power spectrum.  The $\tilde P(k_r)$ spectra were computed for
each simulated image and their average is plotted in
Fig.~\ref{fig:mean_var} using black dots for the unmasked images,
green squares for mask 1 and blue stars for mask 2.

A very small deviation from a perfect power law shape is apparent at
the high frequency end of the recovered power spectra even in the
unmasked case.  This happens because as the filter width approaches
the size of the resolution element, the filter is under-sampled and
fluctuations at those scales are less accurately recovered. There is a
small decrease of power, of not more than 10\% of the expected value,
at k=0.3 pix $^{-1}$ . Below this frequency the recovered power
spectrum is not affected, so if the range of frequencies covered is
large, the high frequency part can be discarded and the spectral slope
can be accurately measured. Similar effects are seen in 1D and 3D.

The high frequency points in Fig.~\ref{fig:mean_var} have the smallest
scatter since they are averaged over many Fourier modes. Therefore,
the statistical uncertainty, discussed in Sec. 6 is small, and can be
smaller than the deviations described above. To avoid assigning too
much weight to these points when fitting the measured power spectra, a
systematic error of 10\% was quadratically added to the estimated
statistical uncertainty, associated with each point.




\begin{table}
\begin{tabular}{|c|cc|cc|cc|c|}
\hline
input&\multicolumn{2}{|c|}{No mask}&\multicolumn{2}{|c|}{Mask
  1}&\multicolumn{2}{|c|}{Mask 2}&norm\\
slope & slope &norm&slope &norm&slope &norm& bias\\
\hline
  0.0&-0.02 & 1.00&-0.03 &1.03&-0.24 &1.33 & 1\\
-1.0&-1.00 &0.92 &-0.99 & 0.94&-1.04 &1.04 & 0.95\\ 
-2.0&-2.00 &0.94 &-2.00 & 0.95&-2.01 &  0.99&1\\
-3.0&-3.01 &1.12 &-2.97 &1.19 & -2.99& 1.24& 1.25\\
-4.0&-4.05 & 1.91&-4.05 & 1.96& -4.00& 2.53& 2\\
\hline
\end{tabular}
\caption{ Results of power law fits to $\tilde P(k_r)$ spectra
    averaged over 100 realizations of Gaussian random fields in 2D
    with and without the mask. The input power spectra were
    normalized to 1 at $k=1$ and have slopes ranging from 0 to -4.
    For the two groups of columns { under the headings Mask 1 and
      Mask 2,} the masks shown in
    Fig. \ref{fig:2Dmask} were applied to the images before calculating
    $\tilde P(k_r)$. For each case, a recovered power law slope and
    normalization are given. Deviations in the normalization are
    largely explained by the normalization bias calculated in
    Appendix~\ref{ap:bias}.}
\label{fitresults}
\end{table}

{ The $\tilde P(k_r)$ spectra were measured for each masked and
 unmasked simulated image and the average of each set was fit with a
 power law model. The input and recovered spectral slopes and
 normalizations, obtained by fitting a power law to the $\tilde
 P(k_r)$ spectra are given in Table \ref{fitresults}.}

The spectral power law slopes are recovered with a deviation not larger
than 0.24 for any slope probed and not larger than 0.05 for slopes of
--1 or steeper. The normalization deviations from unity follow
closely the bias expectation given in Appendix~\ref{ap:bias}, except for
masks similar to Mask 2 and very flat slopes. This mask represents a very
extreme case, containing many small scale gaps that distort the power
spectrum more than larger gaps as is evident in the top panel in
  Fig. \ref{fig:mean_var} and in Table \ref{fitresults}. For simpler
masks and/or steeper power spectra the method recovers the power
spectral parameters accurately and the effect of the mask is
negligible. 

The distortions produced by small gaps covering a large fraction
of the image ($> 50$\%, similar to Mask 2 in Fig.~\ref{fig:2Dmask}) is
not easy to predict as they depend on the gap structure. For
these cases, when the measured power spectrum slope is flatter than
-1, we recommend estimating the distortions through Monte Carlo
simulations using known input power spectra and the same gap
distribution as in the real data.

The accuracy with which power spectral parameters can be recovered
depends on the amount of data available, since this determines the
range of scales covered and the density of independent Fourier
modes, which affects the errors. As an example we fitted the power
spectrum of each individual realization of the images discussed above
to measure the scatter in the recovered power law parameters. In
these fits, the slope and normalization were fitted
simultaneously. The RMS of the recovered parameters are quoted in
Table \ref{individual_fits}. The scatter in the recovered slope
and normalization increase for steeper slopes. The scatter is
slightly larger for masked images, although there is no significant
difference between Mask~1 and Mask~2. Notice that these RMS values
characterize  the scatter around the biased mean values, similar
to those quoted in Table \ref{fitresults}.

\begin{table}
\begin{tabular}{|c|cc|cc|cc|}
\hline
&\multicolumn{2}{|c|}{No mask}&\multicolumn{2}{|c|}{Mask
  1}&\multicolumn{2}{|c|}{Mask 2}\\
input slope& slope &norm &slope&norm&slope &norm \\
\hline
0.0&   0.01& 0.02& 0.02 & 0.03& 0.02 & 0.03 \\
-1.0&  0.01& 0.02& 0.02 & 0.03& 0.02 & 0.03 \\ 
-2.0&  0.02& 0.03& 0.03 & 0.02& 0.02 & 0.02 \\
-3.0&  0.03& 0.05& 0.04 & 0.08& 0.03 & 0.06 \\
-4.0&  0.07& 0.18& 0.07 & 0.20& 0.05 & 0.20 \\
\hline
\end{tabular}
\caption{ Root-mean-square scatter of recovered power law parameters
  for sets of 100 simulated images for each spectral slope and
  mask. The images are (300 pix)$^2$ and Masks~1 and 2 are shown in
  Fig. \ref{fig:2Dmask}. }
\label{individual_fits}
\end{table}

\subsection{Comparison with the $\Delta$-variance method}
\label{sec:comparison}
As mentioned in \S\ref{sec:introduction} and \ref{sec:method_ng} the
method described above is a further development of the
$\Delta$-variance method of \citet{stutzki98, bensch01} and
\citet{ossenkopf08}.  For the periodic data without gaps both methods
produce mathematically equivalent results\footnote{Provided that
  during filtering the images are treated as periodic data sets}.  We
now proceed with a more detailed comparison of the filters performance
for non-periodic data with gaps.

The functional form used in \citet{ossenkopf08}, in the limit
of their parameter $v\rightarrow 1$, is as follows :

\begin{eqnarray}
F_{l,core}(x)&=&\frac{4}{\pi l^2}~ e^{  -\frac{x^2}{\left ( l/2 \right
)^2}} \nonumber \\
F_{l,ann}(x)&=&\frac{4}{\pi l^2}\frac{x^2}{\left ( l/2 \right
)^2} ~ e^{  -\frac{x^2}{\left ( l/2 \right
)^2}},
\label{eq:os}
\end{eqnarray}
where $l$ is the size of the filter.  It is obvious that the
difference between these two filters is the Mexican Hat filter.
However the shape of the second filter $F_{l,\rm ann}(x)$ is very
different from the first one $F_{l,\rm core}(x)$. Therefore it is expected
that the impact of edges and gaps will be different for the filtered
images corrected for these gaps and edges. This is in contrast with the
representation of the Mexican Hat filter as the difference between two
Gaussians with almost equal width, which guarantees that the gaps have
essentially identical impact on both Gaussian convolutions leading to the filtered images.

\begin{figure}
\plotone{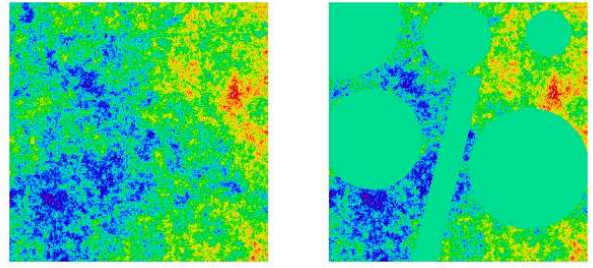}
\caption{ Original image with power spectrum with slope -2 (left) and
  its masked version (right). This non-periodic image was obtained
  by cutting a small section of a larger image. These images are used
  to compare the performance of the two filtering methods.
\label{fig:paos_img}
}
\end{figure}

\begin{figure}
\plotone{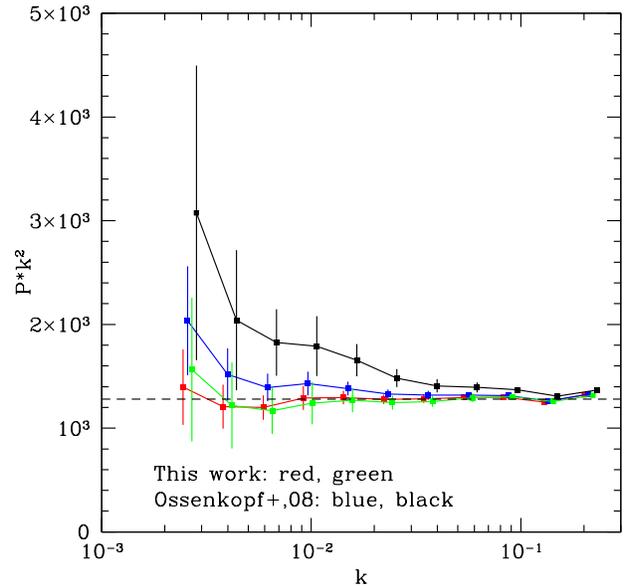}
\caption{ Comparison of the recovered power spectra using the
  difference between two Gaussian and Eq.~\ref{eq:os}. The spectra were
  multiplied by $k^2$, so that the input spectrum is flat in this plot
  (shown by the horizontal dashed line). The spectra were averaged
  over 10 independent realizations.  The red and green points were derived from the original and masked images respectively,
  using the filter described in this work. The blue and black curves
  correspond to the original and masked images respectively, processed
  using the pair of filters from \citet{ossenkopf08}.  For
  non-periodic data sets and/or for data with gaps the filter based on
  two nearly identical Gaussians performs better.
\label{fig:paos_ps}
}
\end{figure}

To verify the above conjecture we made a number of test with
different power spectra and masks. An example of an image and its
masked version is shown in Fig.~\ref{fig:paos_img}. As before the
non-periodic image was obtained by cutting a section of a larger
periodic image, in which a random realization of a Gaussian random field
with slope -2 was generated. The spectra recovered from the original
and masked images for both methods are shown in Fig.~\ref{fig:paos_ps}. The red and green points
were obtained using the procedure described in
\S\ref{sec:method_wg}, for the original and masked cases, respectively. Both for the original and for masked images the
spectrum agrees well with the input spectrum, shown by the horizontal
dashed line. The spectrum, obtained using Eq.~\ref{eq:os} shows larger
deviations from the true input spectrum, especially for the masked
image, shown in black. Similar results were obtained for other types of masks
probed. We therefore conclude that for non-periodic data sets with
gaps the filtering based on two nearly identical Gaussians provides
more robust and accurate results. { We note that the value of $v$ recommended by the authors is $v= 1.5$. Repeating the test for this value
  of $v$ gave similar results to the $v=1$ case and our approach with
  similar Gaussian still proved more accurate.}

\subsection{The 2D sphere: application to CMB analysis}
\label{sec:CMB}
{ 
The same method can be constructed in spherical coordinates. As an
example, in this section we apply the filter formalism to the study of
temperature anisotropies of the Cosmic Microwave Background (CMB)
radiation.  Let
$t(\bmath{n})$ be the CMB temperature in the direction $\bmath{n}$. A
decomposition of the $t(\bmath{n})$ map on the
sphere in a basis of spherical harmonics is,
\begin{equation}
t(\bmath{n} ) = \sum_{l=2}^{\infty}\sum_{m=-l}^l a_{l,m} Y_{l,m}(\bmath{n}
),
\label{eq:almdec1}
\end{equation}
where $Y_{l,m}(\bmath{ n} )$ is the spherical harmonic of order $l,m$
evaluated at the direction $\bmath{ n}$ and $a_{l,m}$ are multipole
coefficients. We assume that $t(\bmath{n})$ is a Gaussian isotropic
and homogeneous random field. The angular power spectrum $C_l$ for
multiple $l$ can be estimated as 
\begin{equation}
C_l = \langle |a_{l,m}|^2\rangle,
\label{eq:Cldec}
\end{equation}
where the averaging is done over $m$. Figure \ref{fig:CMBfit}, shows
the theoretical prediction for the angular power spectrum of the CMB
temperature anisotropies for different angular frequencies (or
multipoles $l$). The units of the y-axis are ${\cal D}_l \equiv
l(l+1)C_l/(2\pi)$.                  

Similarly to the Cartesian 2D case described in \S\ref{sec:method_wg}
one can define a filter as a difference between two Gaussians defined
on a sphere and allow for the data gaps. This yields an image
(cf. eq.~\ref{eq:a9})
\begin{equation}
{t}_{\sigma} (\bmath{n} ) = \left(
\frac{G_{\sigma_1}*t}{G_{\sigma_1}*M}-\frac{G_{\sigma_2}*t}{G_{\sigma_2}*M}
  \right) \times M(\bmath{n} ) ,
\label{eq:fm1}
\end{equation}
dominated by a particular angular scale $\sim \sigma$. As before
the variance of the resulting image is calculated and divided by the
power of the filter. In the
multipole space the corresponding filter has the same properties as
the 2D filter in Cartesian coordinates, described in the Appendix~\ref{ap:power} (see
eq.~\ref{eq:filter}). Namely,
\begin{eqnarray}
\hat{F}_\sigma(l)=e^{-l(l+1)\sigma_1^2/2}-e^{-l(l+1)\sigma_2^2/2}\approx \\
\approx \epsilon l(l+1) \sigma^2 e^{-l(l+1)\sigma^2/2}, 
\label{eq:fl}
\end{eqnarray}
where $\sigma_1=\sigma/\sqrt{1+\epsilon}$, $\sigma_2=\sigma
\sqrt{1+\epsilon}$ and $\epsilon \ll 1$.

The expected shape of the power spectrum can be calculated as:
\begin{equation}
\tilde{C}_{l_r}=\frac{\sum_l C_l |\hat{F}_\sigma(l)|^2 (2l+1)}{\sum_l |\hat{F}_\sigma(l)|^2 (2l+1)},
\end{equation}
where $l_r$ is defined as the multipole where
$|\hat{F}_\sigma(l)|^2$ reaches its maximum.
The corresponding angular power spectrum $\tilde{C}_{l_r}$ is shown in Figure
\ref{fig:CMBfit} with the dashed line. Clearly, the filtering
procedure smears out small scale features, but recovers the overall
shape and normalization of the power spectrum. This smoothed power
spectrum is what we want to recover from the data using the proposed
method.

In real data, the estimation of the angular power spectrum must deal
with the pixelized data and with the presence of the Milky Way and
other contaminants that make necessary the exclusion of some pixels on
the sky map. We used HEALPix\footnote{HEALPix's URL site: {\tt
    http://healpix.jpl.nasa.gov/ }} to deal with the pixelized maps.
The power spectrum is evaluated from the variance of the
filtered images $\displaystyle \langle{t}^2_{\sigma}\rangle$ as
\begin{equation}
\tilde{C}_{l_r} = \frac{\langle {t}^2_{\sigma} \rangle}{\sum_{l}
  \frac{2l+1}{4\pi} |\hat{F}_\sigma(l)|^2 |W_{l,px}|^2}, 
\label{eq:clestm1}
\end{equation}
where $W_{l,px}$ is the window multipole function for the
HEALPix pixel. 

The presence of a mask usually biases the estimation of the power
spectrum multipoles ($C_l$-s) and, at the same time, it couples these
otherwise independent quantities. To test the method we impose a
sky mask that covers $\sim$ 24\% of the sky, including the Galactic
plane and the position of bright radio sources. We also consider two
different hemispheres of data separately, defined by an equatorial
plane perpendicular to the direction (l,b)=(0\degr, 45\degr) in
Galactic coordinates. Therefore, each hemisphere analyzes roughly 38\%
of the sky.  We apply filters, of width of 0.1\degr, 0.12\degr,
  0.22\degr, 0.35\degr,0.42\degr, 0.5\degr, 1\degr, 2\degr, 3\degr,
  4\degr, 8\degr,{ 10\degr and 20\degr}. An example of a masked CMB realization,
filtered at two different scales is shown in Fig. \ref{fig:CMBmap}.
When applying our { thirteen} different filters in the two masked
hemispheres, we obtain $\tilde C_{l_r}$ for each hemisphere, displayed
by the filled green and blue circles, in Fig. \ref{fig:CMBfit}. The
estimates from the two different hemispheres and for each filter scale
$\sigma$ follow closely the theoretical prediction provided by the
dashed line.

For a single realization of the sky, the relative scatter
of power spectrum estimates with respect to the average value is known
to approximately follow the scaling
\begin{equation}
\frac{\Delta \tilde{C}_{l_r}}{\tilde{C}_{l_r}} = \sqrt{\frac{2}{(2l+1)(\Delta l) f_{\rm sky}}},
\label{eq:errscl1}
\end{equation}
which corresponds to the black solid line in the bottom panel of
Fig. \ref{fig:CMBfit}. The symbol $f_{\rm sky}$ denotes the fraction
of the sky covered in each hemisphere (around $\sim$36\% ), and
$\Delta l$ provides the effective width, in multipole space, of each
filter. The blue and green circles show that the estimates of the
angular power spectrum are scattered around the theoretical
expectation by an amount that is not far from the theoretical
prediction. This behavior breaks down at smaller angular scales, due
to effects related to the finite pixel size, which in this case is
slightly below 7 arc minutes.

\begin{figure}
\includegraphics[width=0.48\textwidth]{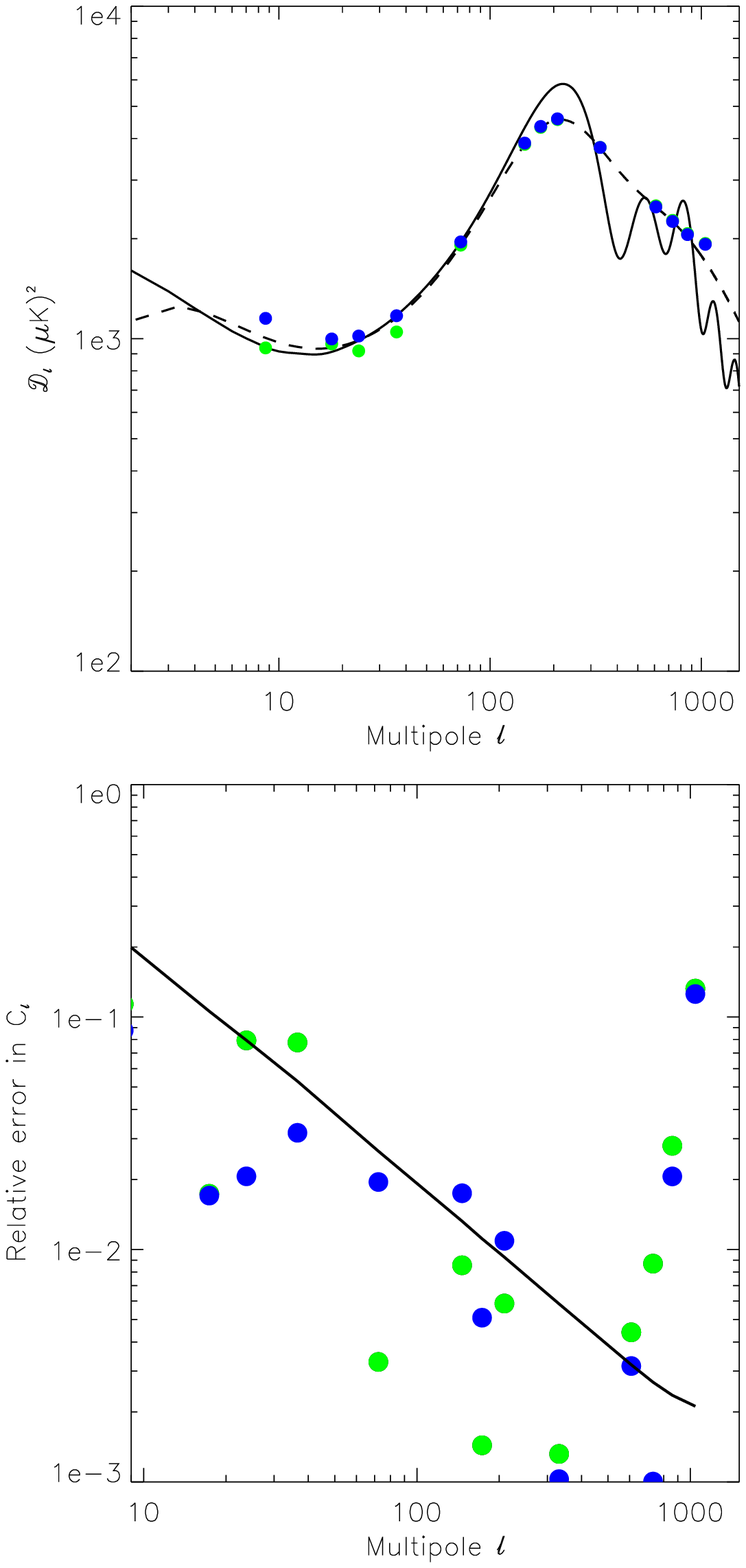}
\caption{Input angular band power spectrum (solid line), the result of
  filtering such power spectrum with scale changing filters (dashed
  line), and the result of computing the angular power spectrum with
  13 different filters in two different hemispheres of a masked single
  CMB sky realization (green and blue circles).  In the bottom panel
  we show the relative error of the power spectrum estimates with
  respect to the theoretical prediction (dashed line in top
  panel). The solid line provides the theoretical expectation,
  $\sqrt{2/(2l+1)/{\Delta_l/f_{\rm sky}}}$ with $f_{\rm sky}=$fraction of the
  sky not being masked. At $l$ values higher than $\sim 800$ the
  filter size reaches the pixel scale and the errors detach from the
  theoretical prediction. }
\label{fig:CMBfit}
\end{figure}

\begin{figure*}
\includegraphics[width=0.49\textwidth]{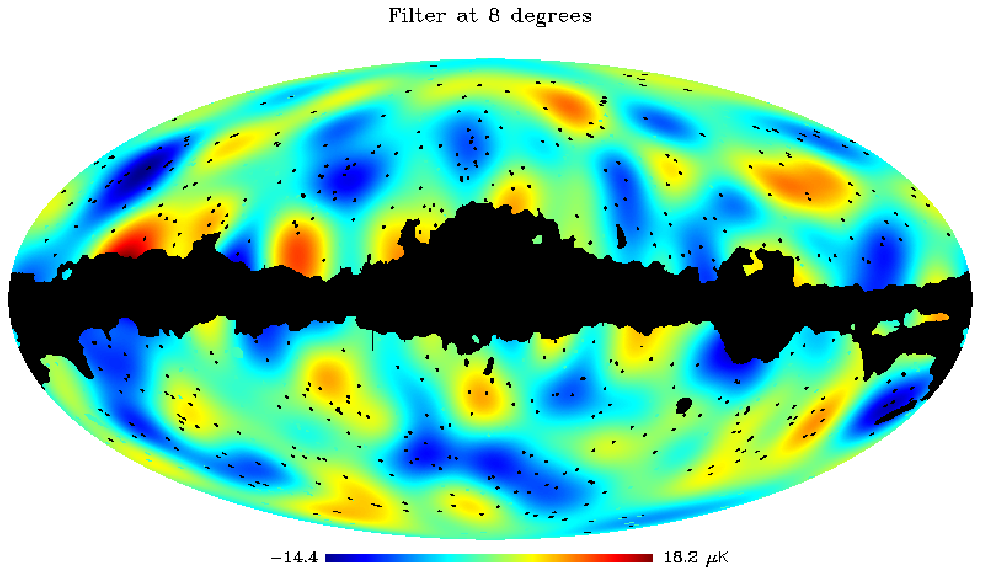}
\includegraphics[width=0.49\textwidth]{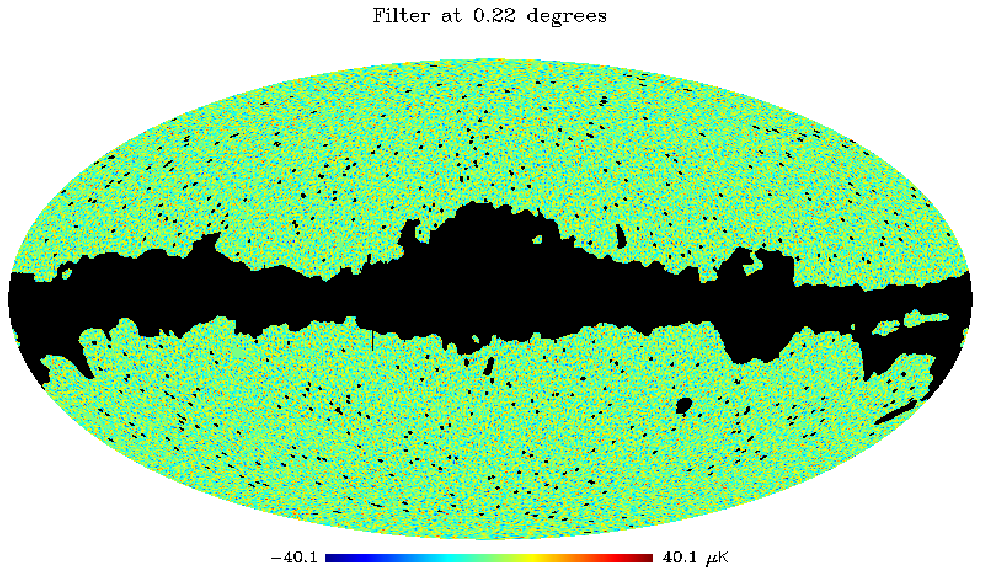}
\caption{Simulated CMB type maps, masked to remove the Galaxy and
  foreground point sources, and then filtered at two different angular
  sales. The maps show no evidence for spurious feature near the edges
  of the mask.}
\label{fig:CMBmap}
\end{figure*}

}

\section{3D data cubes}
\label{sec:3D}

When analysing data of numerical simulations, e.g. a velocity
field in hydrodynamic simulations, one often deals with 3D data cubes. In this section we extend the method on three dimensions and
compare our power spectrum results with calculations using the 
conventional Fourier transform. As before, we generate isotropic Gaussian
random fields with a power law power spectrum
spectrum, in 3D. 
In particular we consider a red noise process with slope $-11/3$,
corresponding to the 3D Kolmogorov power spectrum, which is often
assumed when dealing with the turbulent velocity field.

\begin{figure}
\includegraphics[width=0.48\textwidth]{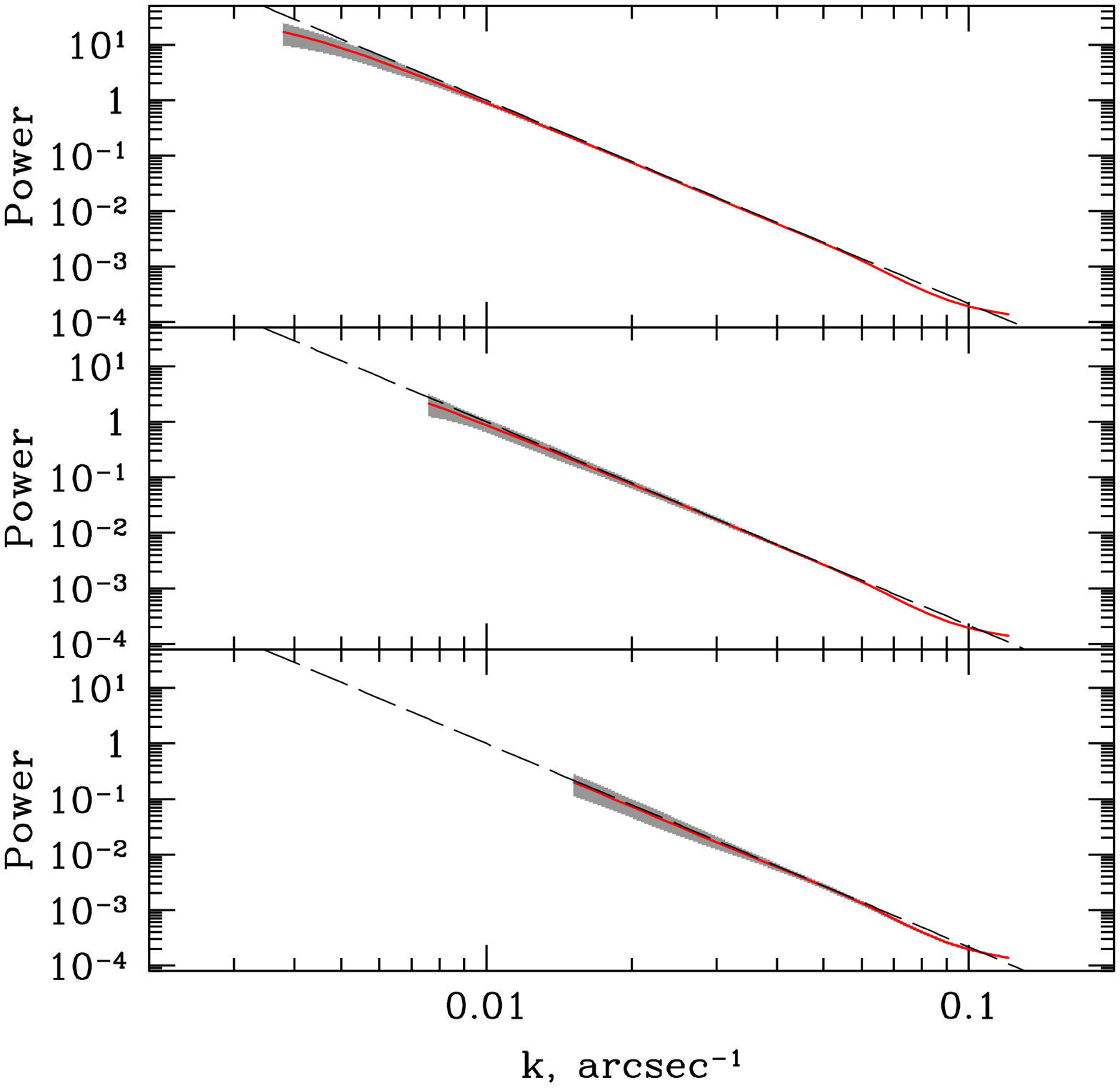}
\caption{ Power spectra calculated through the variance for boxes of
  different sizes (same as Fig. \ref{fig:trim3D}). Top: initial size
  of the box; middle and bottom: $1/2$ and $1/4$ of the original size
  on a side respectively. Red: the mean of 100 cube realization, gray:
  the range of statistical uncertainties { corresponding to 1
    $\sigma$ scatter}.
\label{fig:trimerr3D}
}
\end{figure}

\begin{figure}
\includegraphics[width=0.48\textwidth]{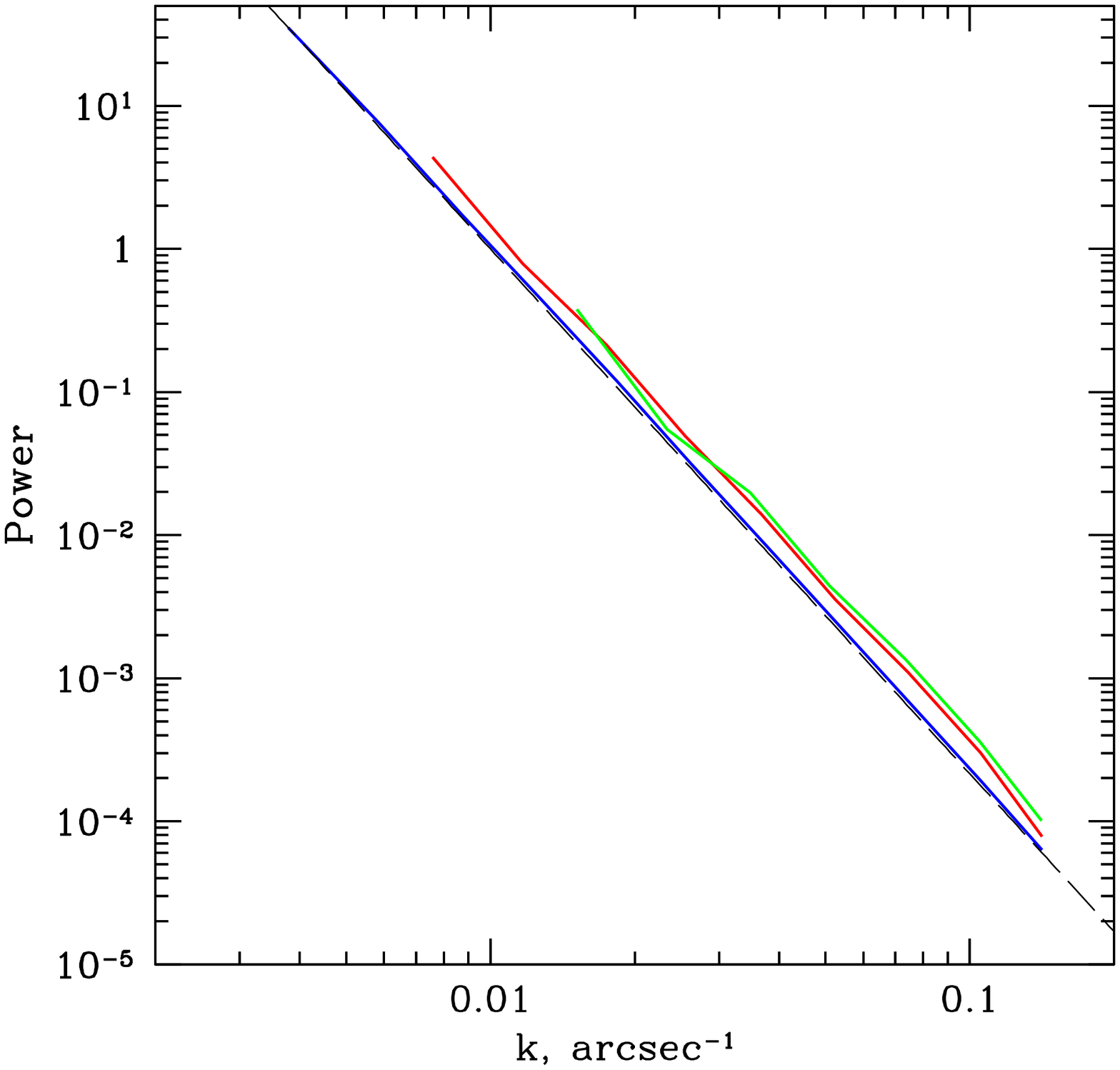}
\caption{{  Fourier power spectrum calculated for a simulated
  box of $64^3$. The power spectrum was obtained by averaging
  power spectra calculated for 100 random realizations with the same slope $\alpha=-11/3$.} The
  input power spectrum is shown with the dashed line. Blue, red and
  green curves show the recovered power spectra from the initial box
  size ($64^3$ cells), and trimmed sections of $1/2$ and $1/4$ of the
  original size on a side, respectively.
\label{fig:trim3D}
}
\end{figure}

The power spectrum $\tilde P(k_r)$ is calculated analogously to the 2D
case, by filtering the cube with 3-dimensional Gaussians. The analysis
in Appendix~\ref{ap:power} is directly applicable for $n=3$. As before we assume isotropy in the random field, so we are
only interested in the power spectrum as a function of spatial scale.

The first question we address is whether the shape of the red noise is
recovered well when the noise extends below the lowest wave numbers set by
the size of the cube. To this end we calculate the power spectrum for
the full simulated box and also for smaller ``trimmed'' boxes, which
are cut from the original larger cube. The mean of 100 power spectra
evaluated through the variance (see \S\ref{sec:method}) for the full
box and trimmed sections, 1/2 and 1/4 of the original size on a
side, are shown in Fig.~\ref{fig:trimerr3D}. The 1$\sigma$ statistical uncertainties for all three cases are shown on
Fig. \ref{fig:trimerr3D} with gray shadows.  The mean of 100 cube
realisations and their trimmed sections are shown by the red lines on
each panel, while the  scatter is represented by the gray
area.  Power spectra calculated
through the conventional Fourier transform are shown in
Fig.~\ref{fig:trim3D}. The input power law power
spectrum is shown by the dashed black line. Since trimmed cubes are
not periodic anymore the power spectrum recovered through Fourier
transform is strongly distorted by leakage of power from very low
frequencies. However, we see good agreement between the input and
recovered power spectra  calculated through the variance
method.  The minor discrepancies are only on the smallest and highest
wave numbers.

Both discrepancies are caused by the fact that the value of the
variance is a convolution of the true PDS with the filter in 
frequency space. As a result at low $k$ the power leaks out if the
full simulated box is used, the simulated cubes effectively
have zero power at frequencies lower then $1/L$. This
effect goes away if a subsection of the original cube is used (see
also 2D case in Fig.~\ref{fig:intro}).

\begin{figure}
\includegraphics[width=0.48\textwidth]{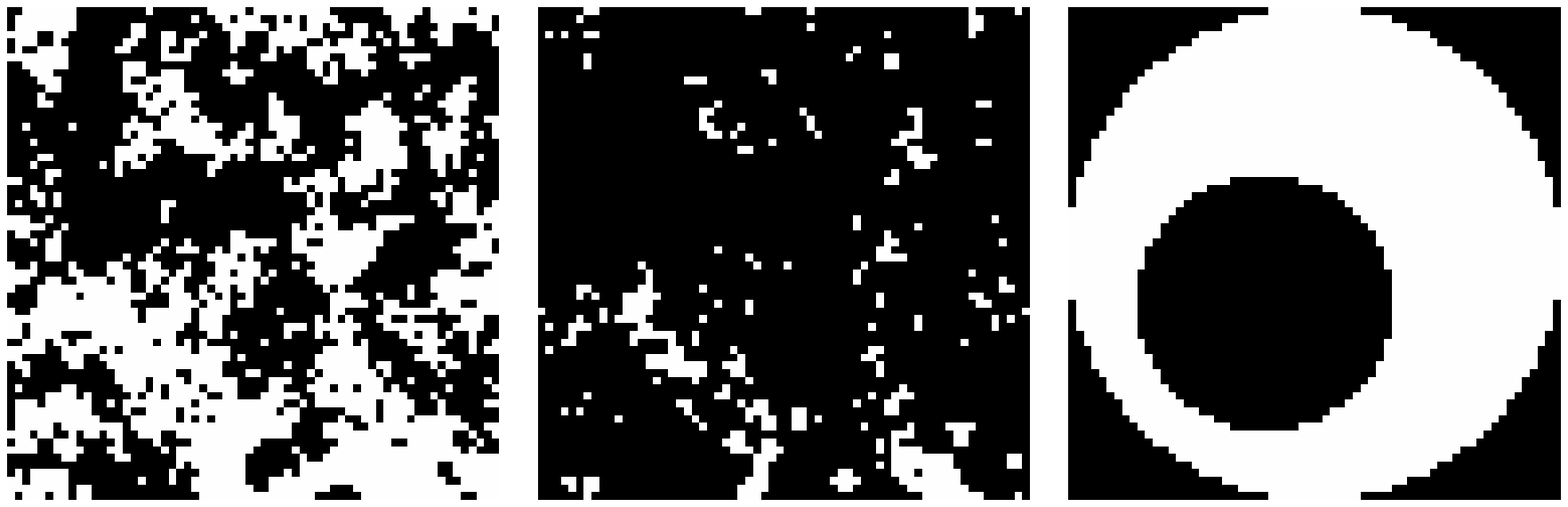}
\caption{ Slices of masks we applied to the data boxes in our
  calculations. Slices are through the center of the box. Black:
  excluded data, $M=0$; white: data used for power spectrum
  calculations, $M=1$. Left and middle panels: masks are generated
  with random gaps. The fraction of discarded data is 50 and 85 per
  cent. Right panel: mask mimics the case of excluded sub-halo.
\label{fig:masks3D}
}
\end{figure}

{ We now proceed by considering the impact of data gaps on the recovered
power spectra. We consider 3D masks covering different fractions of the
cube volume. Slices of masks through the box center are shown in
Fig~\ref{fig:masks3D}. 
The first two masks are generated randomly and the fraction of missing data is
50 per cent in the left panel to 85 per cent in the middle panel. 
Often, when one deals with simulations of galaxy clusters, large
sub-halos or one of the sub-clusters in merging systems are excluded
from the analysis. Consequently, we generated a third 3D mask that
mimics a situation when large sub-halo is excluded. The slice of this
mask is shown on the right panel in Fig. ~\ref{fig:masks3D}.}

\begin{figure}
\includegraphics[width=0.48\textwidth]{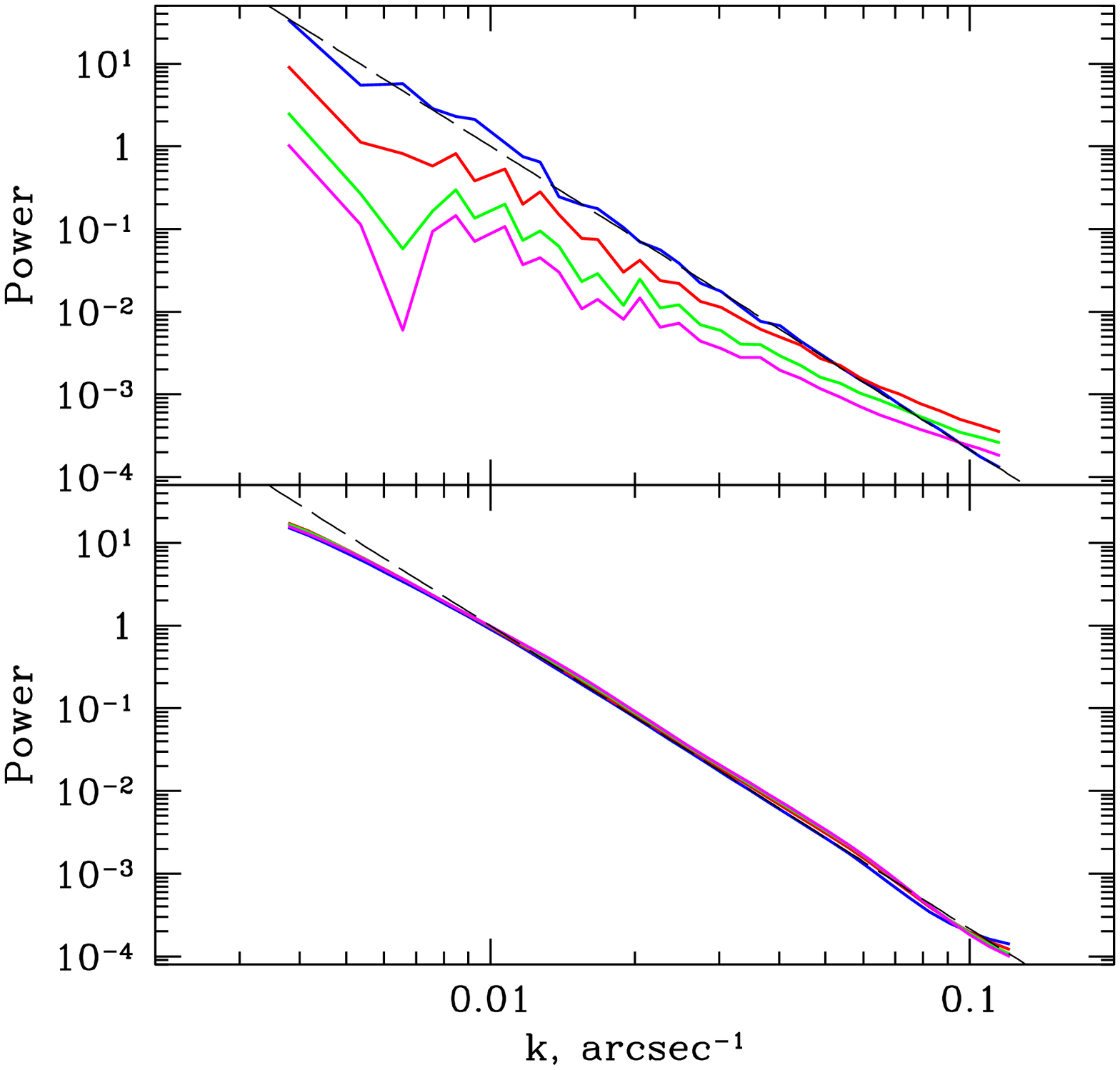}
\caption{ Power spectra calculated for data with gaps (one
  realization). Top: Fourier method, bottom: variance method. Dashed
  curves show initial power spectrum. Blue: 50 per cent of data is
  discarded, see the left panel in Fig. \ref{fig:masks3D}; red: 85 per
  cent of data is excluded, the middle panel in
  Fig. \ref{fig:masks3D}; green: large sub-volume of the box is
  excluded, see the right panel in Fig. \ref{fig:masks3D}. Fourier
  method changes the slope and normalization of the recovered power
  spectrum, while the variance method recovers power spectrum without
  distortions on a large range of wavenumbers.
\label{fig:maskfft3D}
}
\end{figure}

Figure \ref{fig:maskfft3D} shows the power spectrum of the
 data with gaps calculated through the conventional Fourier transform
 and with the variance method. Clearly, the direct Fourier method
 should not be used for data with gaps. The increase of the gaps fraction leads to the
leakage of power causing the flattening of the spectrum and the decrease
of its normalization. At the same time the power spectra
calculated through the variance are perfectly recovered, with only
minor changes on small and large wave numbers even in case when 85 per cent
of the data are in gaps.

\section{1D: Timing analysis}
\label{sec:timing}

A possible 1D application of the method can be found in the study of
light curves of variable objects. The method is of course not
suitable for the search of periodicities, but can be useful for
studies of aperiodic variability, when the broad band shape of
the power spectrum is of interest.

\begin{figure}
\includegraphics[width=0.75\columnwidth,angle=270]{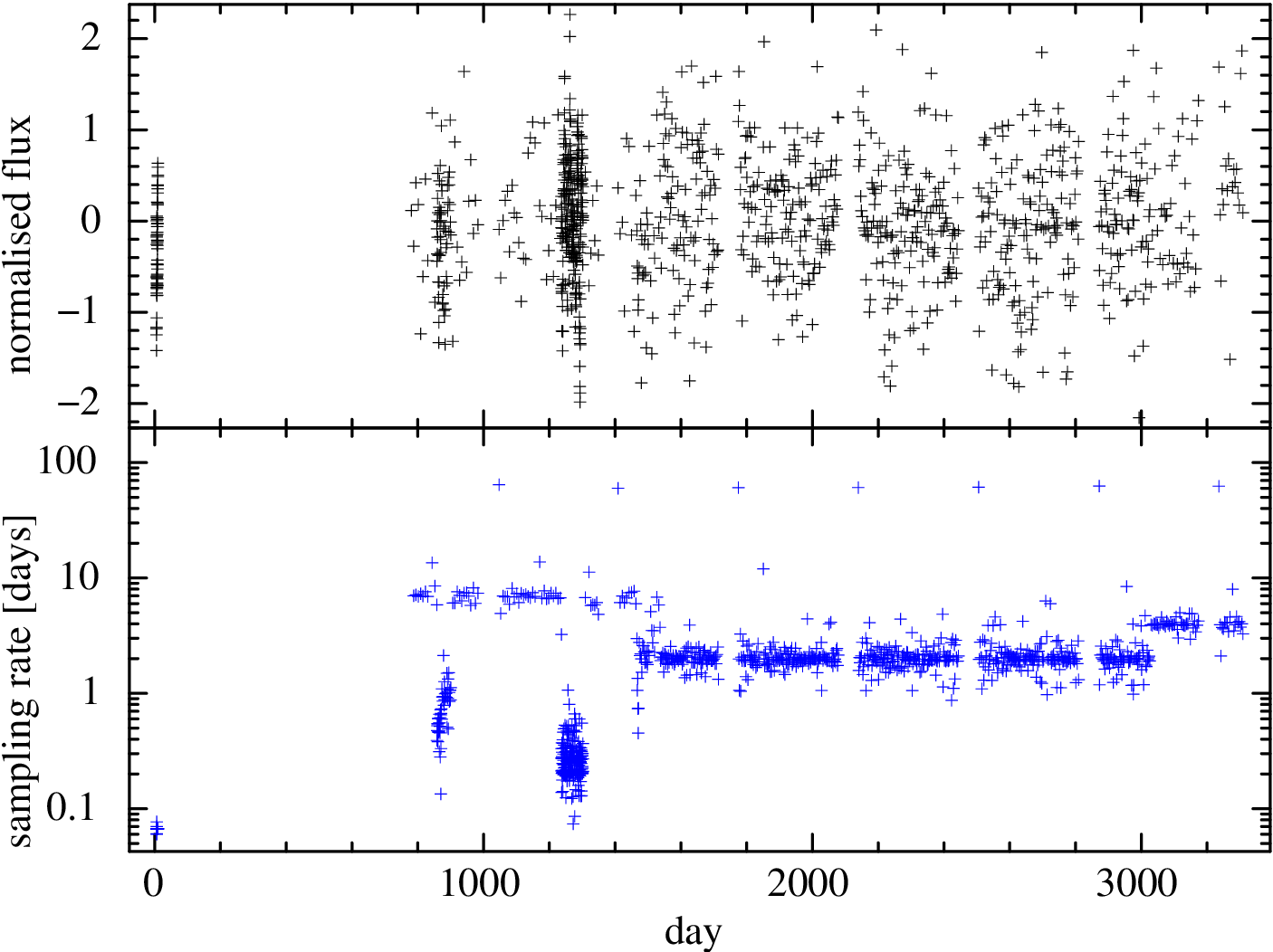}
\caption{ \label{fig:lc}Top: Simulated light curve, typical of AGN long
  term X-ray monitoring, with varying sampling frequency and
  containing yearly gaps. { Bottom: The sampling rate varies with time, here the time difference between consecutive observations is plotted as a function of time, except for the largest gap, after the short, high cadence monitoring. }}
\end{figure}

The top panel in Fig. \ref{fig:lc} shows a simulated light curve typical for long term
X-ray monitoring of an Active Galactic Nucleus (AGN)
\citep[e.g.][]{psresp,markowitz03}.  The power spectra of these light
curves is normally modeled as a single or broken power law, where the
slopes and/or break frequencies are of interest
\citep[e.g.][]{mchardynat,papadakis09}.  These light curves often have
yearly gaps due to visibility constraints and different variability
time-scales are covered by varying the sampling rate. The window
function produced by these gaps causes spurious features in the power spectrum
and different approaches have to be used to
remove these features from the Fourier transform of the light curve
\citep[e.g.][]{psresp,markowitz10,dimitriosSF,Kastendieck}.

{ The case of AGN light curves is essentially different from images and
  simulated data cubes since normally the light curves are not sampled
  continuously over a regular grid, but consist of snapshot
  observations on irregularly spaced time intervals. Long term AGN
  light curves, as described above, suffer strongly from this effect,
  since the timescale coverage is optimized by combining epochs of
  very different sampling rates as shown in the bottom panel of Fig. \ref{fig:lc}. For our example, frequencies from
  $3\times10^{-9}$ to $3\times10^{-5}$ Hz are probed using only 1100 light
  curve points. If it was evenly sampled, this range would require
  20.000 points. This uneven sampling does not pose a problem to the
  variance calculation, since the filter convolution can be performed
  over uneven grids. In this case, the mask is simply the same time
  series, with all flux values replaced by 1. Figure
  \ref{fig:filteredlcs} shows the filtered light curves on different
  timescales produced by this approach, gaps in the time series and
  varying sampling rates do not prevent a clean filtering of the
  fluctuations.

\begin{figure*}
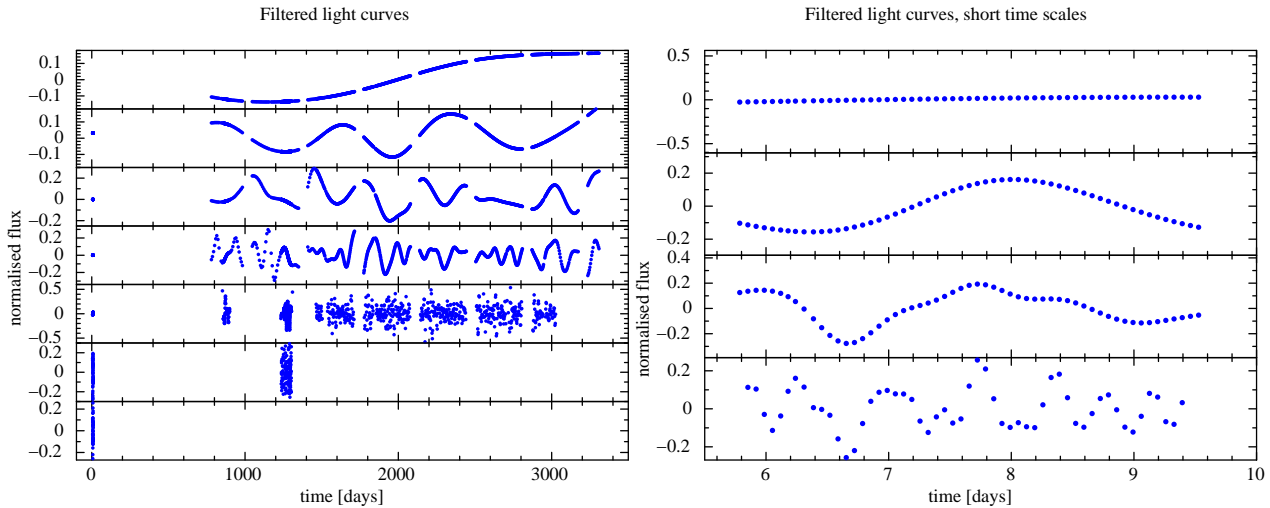

\includegraphics[width=0.8\columnwidth,angle=270]{filteredlcs_fig.ps}
\includegraphics[width=0.8\columnwidth,angle=270]{filteredlcs_short.ps}
\caption{ \label{fig:filteredlcs} Result of the filtering routine on
  the light curve shown in Fig. \ref{fig:lc}. Gaps in the light curve
  do not affect the filtering process. Short timescales are only
  probed by higher cadence light curve segments. This is achieved by
  discarding data points that have less then 6 neighbors within a time
  space of $4\sigma$ for each value of the filter width $\sigma$ while
  calculating the convolution. { The panel on the left shows light
    curves filtered on timescales of 3300, 1100, 366, 121, 13.5, 1.5 and 0.5
    days.} The panel in the right shows a zoom to
  the most intensive monitoring section, at the beginning of the light
  curve, { for filter timescales of 13.5, 4.5, 1.5 and 0.5 days, from top
  to bottom.} }
\end{figure*}

In Figure \ref{fig:filteredlcs} we have only plotted the useful part
of each filtered light curve, these sections are selected to reduce
the effect of aliasing, which is potentially a larger problem for
light curves than for images. In the case of { isotropic fluctuations in
images}, fluctuations on
length scales shorter than the pixel size are averaged out. In the case
of light curves, unless they are continuously exposed, variability on
timescales shorter than the sampling interval adds to that of the
longer fluctuations. This effect aliases power into lower frequency
bands and distorts the measured power spectrum.

Aliasing can be diminished by demanding a minimum number of exposures
within the convolution Gaussian around a given lightcurve point for
this point to count towards the variance calculation. In other words,
if a point in the light curve has no neighbors closer than $4\sigma$,
for a given filter width $\sigma$, then it cannot provide information
on the corresponding frequency $k_r$. Therefore, although the point
can participate in the convolution of the light curve, it should not
be counted in the variance for that frequency. The procedure applied
on the simulated light curve counts the number of neighbors for each
point in the light curve while making the convolution and later
discards points that have less than 6 neighbors for a given filter
width, producing the filtered lightcurves in Figure
\ref{fig:filteredlcs}. Therefore, the shorter timescales are only
probed by the more intensively sampled sections.

We tested the recovery of the input power spectra for unevenly sampled
light curves for different minimum numbers of neighbors within
$4\sigma$ of each point. Figure \ref{fig:pdslc_sigma4} shows an
example of these tests, for simulated light curves of power law slope
-1. Each power spectrum is the average of 100 simulations, to average
out the fluctuations between different realizations. In the Figure,
the black solid line represents the power spectrum with no minimum
number of neighbors required, so that all light curve points count
towards all frequency bins. The peak in the middle is the effect of
aliasing and the drop at high frequencies is the result of counting
many isolated points in the variance, since these have all the same
value they reduce the variance artificially. The red dashed line uses
at least two neighbors and already corrects this last problem, while
the aliasing is not resolved. Requiring 6 neighbors (pink dot-dashed
line) already solves a large part of both problems, while requiring 10
neighbors (blue dotted line) produces a power spectrum quite close to
the input power law. Using more neighbors does not improve the
recovered power spectrum but it reduces the probed frequency range,
since at high frequencies the sampling rate itself limits the number
of possible neighbors.

\begin{figure}
 \includegraphics[width=7cm,angle=270]{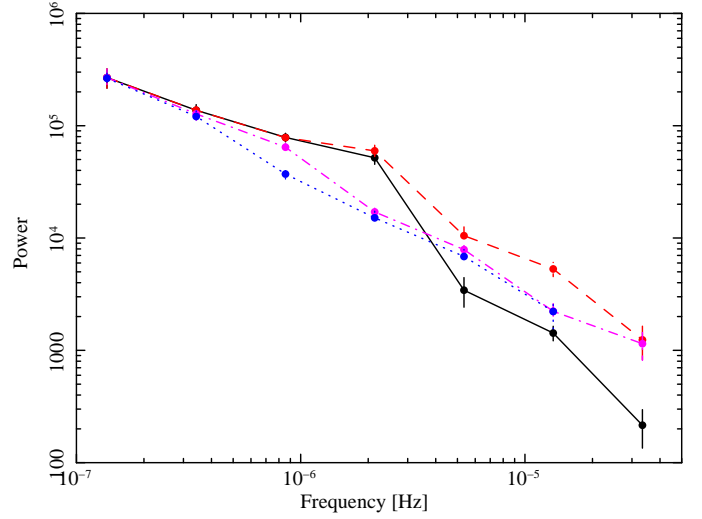}
\caption{ \label{fig:pdslc_sigma4}  The minimum number of neighbor
  points required to count a point in the variance calculation affects
  the shape of the resulting power spectrum. The power spectra above
  are each an average of 100 realizations, requiring 0 neighbors
  (solid, black), two neighbors (red, dashed), six neighbors (pink ,
  dot-dashed) and ten neighbors (blue, dotted). Requiring more than 10
  neighbors does not improve the recovered power spectrum and reduces
  the sampled frequency range as explained in the text. Error bars
  denote the RMS scatter in each set of simulations.}
\end{figure}

Finally, the entire power spectrum of the full range of frequencies covered by the light curve in Fig. \ref{fig:lc} can be computed directly. Figure \ref{fig:pdslc} shows the result of applying the filter formalism using a minimum of 6 neighbors to simulated light curves with the same sampling pattern and a bending power law power spectrum. For this Figure, 100 random realizations were averaged together. At low frequencies the flat input slope and extreme mask produces the slope bias expected from the results in Table 1 while the high frequency normalization is biased upwards by a factor of 1.3, as expected from Appendix~\ref{ap:bias}. The approximate slopes, break and normalization of the input power spectrum, shown by the solid line, are clearly recovered. }

\begin{figure}
\includegraphics[width=6cm,angle=270]{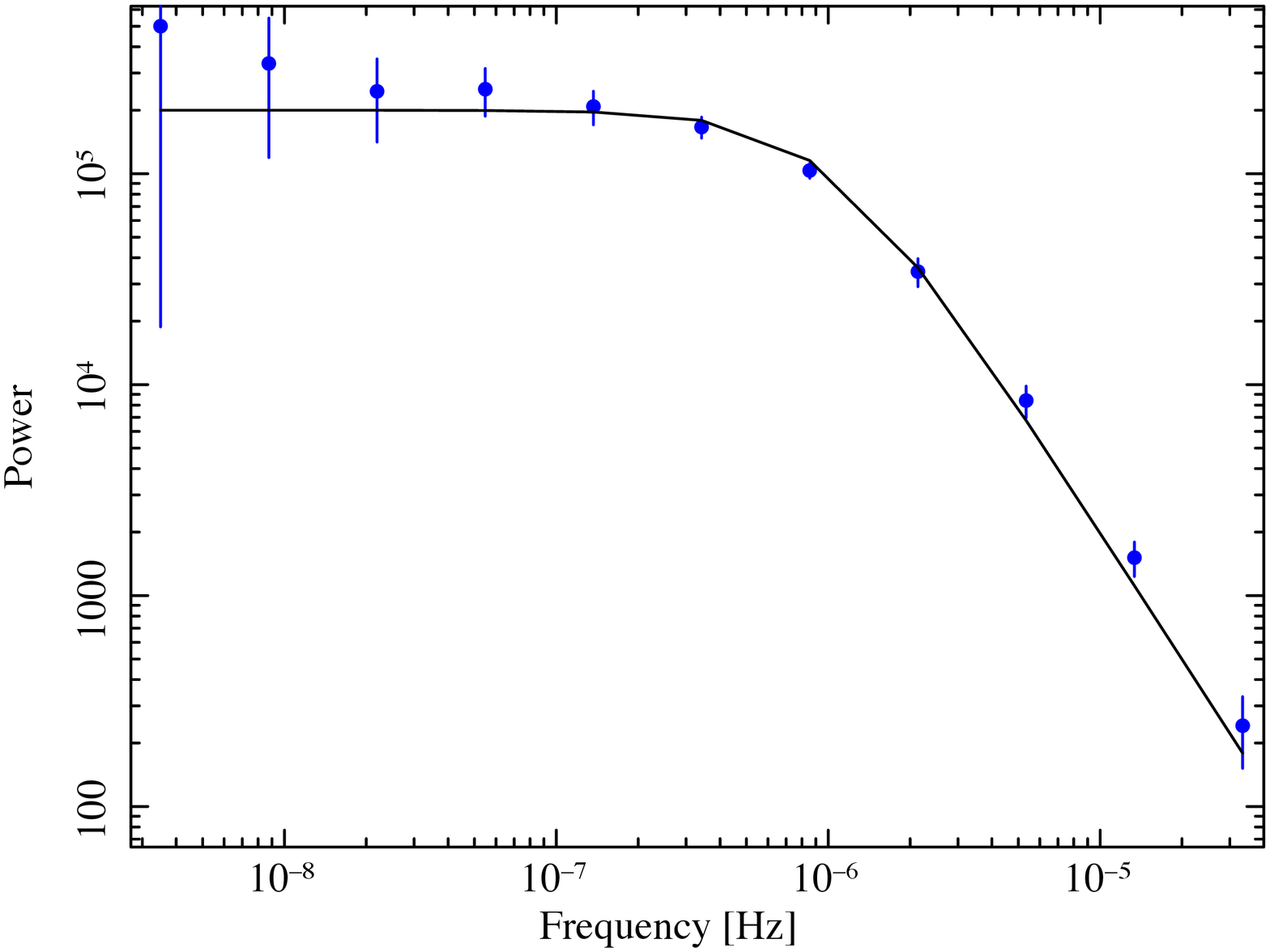}
\caption{ \label{fig:pdslc}  Power spectrum of the simulated light
  curve with the same sampling as in in Fig. \ref{fig:lc} with a flat
  input power spectrum which bends to a power law slope of --2 at a
  frequency $10^{-6}$ Hz. The black symbols correspond to $\tilde
  P(k_r)$ averaged over 100 realizations of the light curve. Error bars denote the RMS scatter in each set of simulations.}
\end{figure}

For light curves similar to the one shown in Fig. \ref{fig:lc}
the power spectrum is usually estimated through Monte Carlo
simulations of sample light curves with given power spectrum
parameters, attempting to recover only power law slopes, breaks and
normalizations. 

The simplicity of the variance method is that it can be applied
directly on the full light curve, regardless of the gap and sampling
distributions and reproduce the overall shape with approximately
correct normalization of the power spectrum. This will be particularly
useful for large samples of AGN light curves from large-area repeated
surveys, where estimation of aliasing and red noise leak for each
particular AGN becomes impractical.  For comparison, power spectrum
fitting of the light curve in Fig. \ref{fig:lc} using PSRESP
\citep{psresp} takes
several hours whereas $\tilde P(k_r)$ and its error estimate can be
obtained in a few seconds on the same computer and both methods
recover the spectrum with similar accuracy.

\section{Error estimation}
\label{sec:error}

As with other methods, it is possible to predict the uncertainty of
the power estimates by assuming certain properties of the underlying
variability process. In this case we will assume that the data being
treated correspond to a Gaussian field, so that independent points in
the power spectrum are distributed around their mean values as a
$\chi^2$ distribution with two degrees of freedom, with standard deviation equal
to the mean.  This means that the uncertainty of the power for
  each individual Fourier mode is equal to the power
  itself. Therefore, to calculate the expected uncertainty in a
  frequency bin, it is only necessary to combine quadratically the
  power at the frequencies that fall within the bin.

 The effect of the filter in frequency domain is to make a weighted
 average of the power spectrum estimates within the width of the
 filter. Therefore, the expected uncertainty, $E$, of $\tilde P(k_r)$
 can be calculated as a weighted mean of the $P(k)$ values that fall
 under the frequency filter, where the values are combined
 quadratically.

Independent Fourier modes in each dimension are separated in frequency
by $1/N_i$ where $N_i$ is number of pixels in dimension $i$. This
spacing needs to be taken into account when calculating the
uncertainty in order to include the correct number of power estimates
under the filter. In the case of isotropic fluctuations in $n$
dimensions, where we only calculate the power as a function of
$k=\sqrt{\sum_{i=1}^{n}(j_i/N_i)^2}$, different groups of indices
producing the same value of $k$ are averaged together and this should
also be considered in the uncertainty calculation.

The correct density of Fourier modes is easily obtained by summing
over all the independent modes, or equivalently over half the data points, i.e. over points $j$, from 1 to $N_1/2$ in
the first dimension and from 1 to $N_i$ in all the following
dimensions as

\begin{equation}
E(k_r)=\frac{\sqrt{\sum_{j=1}^{N/2}(P(j/N)\hat
    F^2_{k_r}(j/N))^2}}{\sum_{j=1}^{N/2}\hat F^2_{k_r}(j/N)}.\label{eq:error}
\end{equation}
for 1 dimension and 
\begin{equation}
E(k_r)=\frac{\sqrt{\sum_{j=1}^{N_1/2}\sum_{l=1}^{N_2}(P(k)\hat
    F^2_{k_r}(k))^2}}{\sum_{j=1}^{N_1/2}\sum_{l=1}^{N_2}\hat F^2_{k_r}(k)}.\label{eq:error}
\end{equation}
for 2 dimensions. For larger $n$ the formula is similar.

The filter $\hat F_{k_r}(k)$ is centered on $k_r$ in frequency domain
and is given in equation A6. Clearly for larger data sets (larger
$N_i$), the independent modes are spaced more closely in frequency so
more power spectrum points are averaged together under the same filter
width and the uncertainty decreases accordingly as
$1/\sqrt{\Pi_{i=1}^{n} N_i}$. Since the number of Fourier modes within
the filter width grows linearly with increasing frequency in each
dimension, the
expected error spectrum is steeper than the powerlaw power spectrum by 0.5 in
1-D, by 1 in 2-D and by 1.5 in 3-D. As shown in the next section, this
prediction is well matched by the data.

 The expected error can be calculated in one pass through the data
points per frequency $k_r$, so it does not imply a significant
computational effort.

The assumption of a Gaussian field implies that each independent
frequency will have its power distributed as a $\chi^2$ distribution
with two degrees of freedom but the average over many such points will
lead to a Gaussian distribution. For low frequencies the number of
Fourier modes within the filter is small, so the distribution of
powers deviates noticeably from Gaussian. However, since the filter is
broad in frequency, in 1D $k_r=4/T$ already contains enough points to
produce an approximately symmetric distribution of measured power. In
2D and 3D cases the symmetry is achieved at even lower frequencies since
the number of Fourier modes within the filter grows as $k^{n}$.

The values of $P(k)$ in the error formula can be estimated by
interpolating between measurements of $\tilde P(k_r)$, given that the
true underlying form is not known a priori. 
Using the measured power spectrum in the error formula
automatically takes into account the normalization bias and any other
distortions, so that the resulting errors always represent the
statistical uncertainty around the measured values. If the spectrum is
finally rescaled by the normalization bias, the errors should be
rescaled accordingly.

{
\subsection{Comparing expected error with measured scatter}

The error formula predicts the  scatter of $\tilde P(k_r)$
measurements around the underlying power spectrum for a Gaussian noise
process. We tested the accuracy of this approach by comparing directly
the predicted error to the RMS scatter in 100 realizations of
a noise process. Figure \ref{fig:var_error} shows this comparison of
measured scatter in markers and predicted errors in lines, both
relative to the measured power spectrum. Solid
lines and nearby markers correspond to 1D time series without gaps
with input power spectra slopes of -1 and -2 (the power spectra
themselves are not shown here). There is a very good agreement between
the measured RMS spread and the estimated uncertainty $E$ in the case
of continuous data sets.  } The error estimate described in this
section is equivalent to the expected scatter from Fourier analysis in
the case of continuous data sets. The analysis above shows that the
scatter around the mean power in binned Fourier transforms and in
$\tilde P(k_r)$ spectra are identical, so both methods are equivalent
when measuring low resolution power spectra of data without gaps.

\subsection{Effect of gaps on the error estimate}
Evidently, gaps in the data will reduce the number of available
independent points and so the uncertainty in the power spectrum
measurement should increase. As a first step to model this effect we
scale $E$ by the number of points actually available in the data
($N_a$) compared to the total number of points if there were no gaps
$N$: $E'=E\times \sqrt{N/N_a}$.

\begin{figure}
\includegraphics[height=0.85\columnwidth,angle=270]{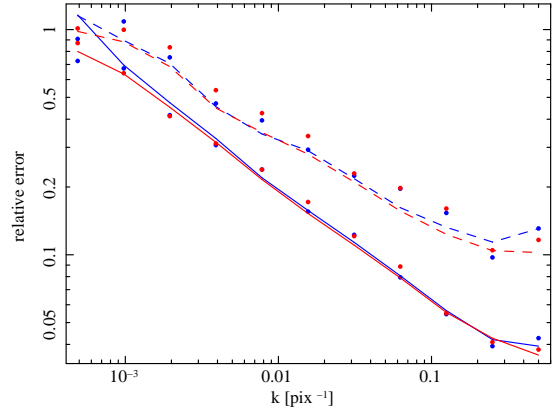}
\caption{ \label{fig:var_error}Comparison of predicted relative error (lines)
  with measured RMS scatter (markers), for sets of 100 $\tilde P(k)$ of
  simulated light curves. The solid lines and corresponding markers are calculated
  for continuous light curves of 2048 points and power law slopes of
  --1 (blue lines and markers) and --2 (red lines and markers). The dashed lines and markers show the predicted and
  measured errors for the same light curves after 85\% of the data
  points have been masked out producing gaps of random length. For this
  range of slopes at least the analytic prediction of the uncertainty
in Eq. \ref{eq:error} works well. }
\end{figure}

{ As a further refinement, it is convenient to count the number of
  available points as a function of $k_r$, since small gaps do not
  affect the power estimates at low wavenumbers. This estimate is
  shown by the dashed lines in Fig. \ref{fig:var_error}. A good match
  to the measured scatter (markers closest to the dashed lines) was
  obtained when points in gaps were discounted only if the gap was
  longer than 10\% of the filter spatial scale ($1/k_r$). The errors
  and scatter measurements correspond to the same sets of simulated
  light curves discussed above, this time masked by gaps of random
  lengths so that 85\% of the points were discarded.} The uncertainty
for each slope increases and the increase is larger at small spatial
scales, which are affected by more gaps. In any case, the increase in
uncertainty for any length scale is between $E$ and
$E\times\sqrt{N/N_a}$, so that for a small fraction of missing points
the dependence of the error increase on $k_r$ can be neglected.

 In the case of light curves with uneven sampling and very
  different sampling rates it is more convenient to scale the error by
  the lengths of useful data stretches to the total length. As before,
  useful light curve stretches can be selected for each frequency
  $k_r$ by requiring a minimum number of neighbors for useful data
  points. All useful segments are added to produce $T_u(k_r)$ and the
  error is scaled as $E\times\sqrt{T_{\rm tot}/T_u}$, where $T_{\rm
    tot}$ is the total length, used to calculate the frequency
  spacing.  

When using these error estimates to fit models to the power spectrum
it is necessary to use only independent measurements that, given the
mixing effect of the filter, should be separated by at least a factor
or 2 in frequency.

\begin{figure}
\includegraphics[height=8cm,angle=270]{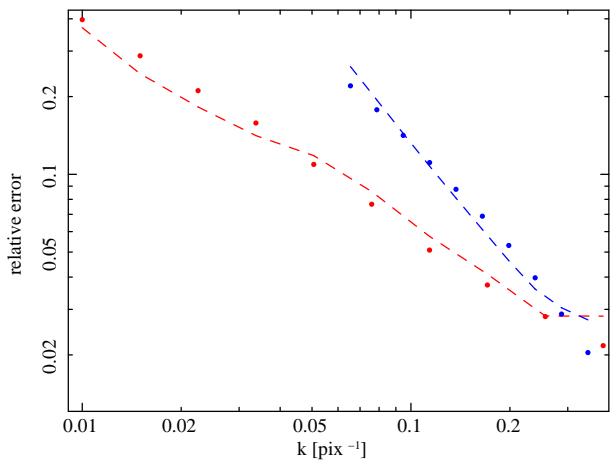}
\caption{\label{examples}{ Relative error prediction for 2D (red) and 3D
  (blue) data. In this example the input power spectrum of the 2D
  data is a broken power law of low frequency slope --0.5 and high
  frequency slope --2.5, and for the 3D data is a power law of slope
  --11/3. The images have been masked out by random sized gaps
  covering 50\% of the image. The predicted errors divided by the
  average power spectrum are shown by the
  dashed lines and the measured RMS scatter of the 100 realizations is
  shown by the blue and red markers. The error formula in Eq. \ref{eq:error}
  predicts the errors accurately for simple and broken power law
  spectra in images and data cubes as well.}}
\end{figure}

The analytic error estimate was compared to Monte Carlo simulations of
Gaussian fields in 1, 2 and 3D, for simple power law or broken power
law power spectra of zero or negative slopes. Although this is a
limited subset of all possible power spectra it does cover a wide
range of astrophysical phenomena of interest.  More examples are given
in Fig.~\ref{examples}. Here 2D images with broken power law power
spectra and affected heavily gaps as shown in mask 2 of
Fig.~\ref{fig:2Dmask} are examined. The decrease in number of points
was estimated from gap size distribution in the x direction only to
estimate the errors. Since the gap structure is largely isotropic,
this simple 1D estimate gave a good approximation, shown by the { dashed}
line, error estimates from RMS scatter of 100 trials is shown in black
markers. In the same figure we plot the predicted error and RMS
scatter of 100 simulated data cubes with power law slope --11/3, also
masking out 50\% of the data points. The mask in this case is similar
to that shown in Fig.~\ref{fig:masks3D}, which affects all scales in
approximately the same way so the effect of the mask can be estimated
simply by rescaling the predicted error by the square root of the
ratio of the total number of points to the number of points actually
available.

\subsection{Data with periodic gaps}

\begin{figure}
\includegraphics[width=6.5cm,angle=270,bb=50 20 600 700, clip]{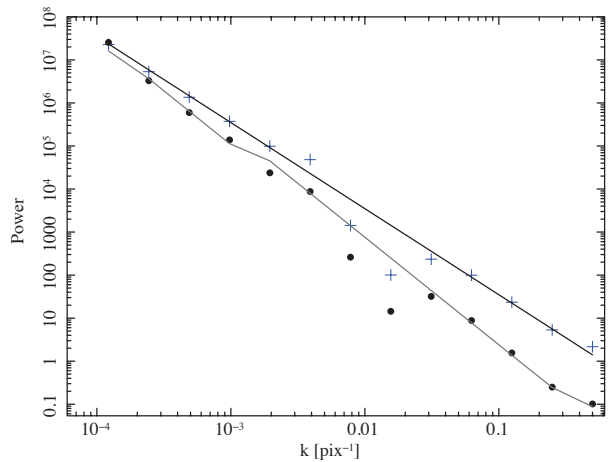}
\caption{\label{fig:periodic}Effect of periodic gaps in 1-D data. The
  above $\tilde P(k_r)$ spectrum is the average of 100 realizations of
  slope --2 light curves masked by periodic gaps, where every 20 data
  points are followed by 80 gap points. The RMS spread of the
  simulations is shown by the black markers and the uncertainty
  predicted by Eq. \ref{eq:error} is shown by the corresponding black line. The
  average power spectrum in blue crosses fits well the input power
  spectrum shown by the top black line and the error estimate works well
  for frequencies above $(2\times T_d)^{-1}=0.025$ and below
  $(4\times(T_d+T_g))^{-1}=0.0025$, the power spectrum for frequencies
  in between these values should be discarded. }
\end{figure}

A special case that we considered is largely found in 1-D light curves
where observational constraints produce strictly periodic gaps. If the
data and gap stretches are of equal length or the data stretch is
longer, then the variance method and error estimate can be directly
applied. However, if the gaps ($T_g$) are longer than the data
stretches ($T_d$) and the sampling pattern is strictly periodic, then
some timescales are effectively not probed by the data and the method
cannot be expected to produce reliable power spectral estimates for
the corresponding frequencies. Frequencies above $(2\times T_d)^{-1}$
and below $(4\times(T_d+T_g))^{-1}$ are properly reproduced and the
error estimate applies, as shown with an example in
Fig.~\ref{fig:periodic} . The power spectrum of frequencies within
this range are not reliable and should be discarded.

Unsurprisingly, it is possible to recover the power spectrum for the
timescales that are well sampled, either shorter than each individual
data stretch $T_d$ or longer than a few times the sampling
interval. In cases like this it is common to make separate estimates
for the high and low frequency power spectrum, by averaging together
individual power spectra from short stretches for the first case and
by computing the power spectrum of low resolution light curves,
averaging together data over $T_d$ and using the sampling interval as
time step for the latter. One of the advantages of the variance method
is that it can compute reliably the same range of frequencies directly
from the complete light curve, without any further manipulation of the
data.

\section{Conclusion}
\label{sec:conclusion}
A simple method for estimating a low resolution power spectrum from
data with gaps is described. Essentially the variance associated with
a given scale is calculated by convolving the image with a Mexican Hat
filter. Gaps in the data, described by the mask $M$ are corrected
for by representing the Mexican Hat filter as a difference between two
Gaussian filters, convolving the image and mask with these filters and
dividing results before calculation of the final Mexican Hat-filtered
image. The variance of the filtered image is then calculated and the
power spectrum is evaluated by repeating the procedure for different
filter scales. The calculated power spectrum is smeared out by the
width of the filter, so sharp features are lost, but the broadband
spectral shape and normalization are recovered well.

The strength of the method is that it is simple and robust. It can
deal with severe gaps in the data and produce accurate power-spectral
power law slopes and normalizations with no tuning parameters and is
computationally cheap. By dividing the filtered images by their
correspondingly filtered masks and maintaining the calculation in the
space domain, the method cleanly compensates for data gaps even if
these have complicated shapes and cover significant part of the data
set. We use simulations to show that the power spectrum recovered from
complete data sets and from their masked versions are consistent and
no additional biases are introduced by the masks.  The method can be
applied straightforwardly to 1D timing analysis, 2D imaging of, for
example, brightness fluctuations in galaxy clusters, and higher
dimensional data cubes from numerical simulations.

\section*{Acknowledgments}
We are grateful to the anonymous referee for many useful comments and
suggestions which helped to improve this paper. We also thank
A. Banday for helpful discussion on CMB data analysis. PA acknowledges
support from Fondecyt grant number 11100449. IZ would like to thank
the International Max Planck Research School (IMPRS) in Garching. This
work was partly supported by the Division of Physical Sciences of the
RAS (program OFN-17).  

\appendix

\section{Cartesian coordinates}
\label{ap:power}
Consider a $n$-dimensional `image' $I(x)$, where $x$ is
$n$-dimensional vector, and corresponding mask $M(x)$ with values of
either 0 or 1.  $M(x)=0$ means that this particular region of the
image does not contain useful information and should be ignored when
calculating the power. We first consider Cartesian coordinates. Our
goal is to estimate typical amplitude of the power spectrum for a
given spatial scale $a$ or, equivalently, given wave number
$k_r$. Here and below we adopt the relation $k_r=1/a$ without a factor
$2\pi$. The image is assumed to be an isotropic Gaussian random field
so we will only compute the power as a function of the scalar $k_r$.

We start by defining a Gaussian filter in spatial domain, which will
be convolved with the image:
\begin{equation}
G_{\sigma}(x)=\frac{1}{(2\pi\sigma^2)^{n/2}}e^{-\frac{x^2}{2\sigma^2}}.
\end{equation}
Assuming first that the mask is equal to 1 everywhere we can combine two
Gaussians to create a filter which selects fluctuations with a given
scale (see Fig.~\ref{fig:filter}):  
\begin{equation}
F_{k_r}(x)=G_{\sigma 1}(x)-G_{\sigma 2}(x),
\label{eq:fs}
\end{equation}
where $\sigma_1=\sigma/\sqrt{1+\epsilon}$, $\sigma_2=\sigma
\sqrt{1+\epsilon}$ and $\epsilon << 1$. The shape of the filter is
identical to the Mexican Hat in the limit of small $\epsilon$. In
Fourier space the corresponding filter is:
\begin{equation}
\hat{F}_{k_r}(k)=e^{-2\pi^2 k^2 \sigma_1^2}-e^{-2\pi^2 k^2 \sigma_2^2}.
\end{equation}
Making a Taylor expansion for $\epsilon$ we obtain  
\begin{equation}
\hat{F}_{k_r}(k)\approx \epsilon 4\pi^2 k^2 \sigma^2  e^{-2\pi^2 k^2 \sigma^2}.
\end{equation}
The filter shape is therefore independent of the value of $\epsilon$
in the limit of small values so that the Taylor expansion is
valid. The value of $\epsilon$ only determines the normalization of
the filter and this effect is canceled in the conversion of { the variance of the filtered
image, $V_{k_r}$}
to $\tilde P(k_r)$ below. The peak of the filter is at $\displaystyle
k=\frac{1}{\sqrt{2\pi^2}}\frac{1}{\sigma}$ (see
Fig.~\ref{fig:filter}). It is therefore natural to relate the
characteristic scale $k_r$ corresponding to this particular filter
width $\sigma$ by:
\begin{equation}
\sigma=\frac{1}{\sqrt{2\pi^2}}\frac{1}{k_r}\approx \frac{0.225079}{k_r}.
\end{equation}
Thus the final expression for the filter in Fourier domain is: 
\begin{equation}
\hat{F}_{k_r}(k)\approx 2 \epsilon \left (\frac{k}{k_r}\right )^2 e^{-\left (\frac{k}{k_r}\right )^2}.
\label{eq:filter}
\end{equation}
Convolving the image $I(x)$ with the filter $F_{k_r}(x)$ described by
eq. \ref{eq:fs} and integrating the square of the convolved image is
equivalent to integrating the product of the power spectrum $P(k)$ and
the square of the Fourier transform of the filter,
$\hat{F}_{k_r}(k)$. Below, $V_{k_r}$ is the variance of the filtered
image $(F*I)$ since the mean of this filtered image is zero.
\begin{eqnarray}
V_{k_r}=\int(F*I)^2 d^nx=\nonumber \\
=\int P(k)|\hat{F}_{k_r}(k)|^2 d^nk. 
\label{eq:van}
\end{eqnarray}
Notice that in eq.~\ref{eq:filter} the filter drops to zero for $k$
larger and smaller than $k_r$ so that the function $P(k)$ can be
approximated by $\tilde P(k_r)$ and taken out of the integral:
\begin{eqnarray}
V_{k_r}\approx  4 \epsilon^2 P(k_r) 
\int \left (\frac{k}{k_r}\right )^4 e^{-2\left (\frac{k}{k_r}\right
  )^2} d^nk= \nonumber \\ 
=\epsilon^2 P(k_r) n\left ( \frac{n}{2}+1 \right) 2^{-\frac{n}{2}-1 }
\pi^{\frac{n}{2}} k_r^n
\label{eq:expected}
\end{eqnarray}
The last approximation breaks down for very steep power spectra, of
positive or negative slope, where its product with the filter does not
drop rapidly to zero for values of $k$ different to $k_r$.

The quantity $V_{k_r}$ is calculated from the image by following the
filter and difference method described and expression
\ref{eq:expected} relates this quantity to the power $P(k)$ through a
scale-dependent normalization. Therefore, the power $\tilde P(k_r)$
can be evaluated.

In practice the mask is not unity everywhere. In this case the above
calculation is done by convolving the image and mask with the two
Gaussian filters, dividing the convolved images; subtracting results
and applying the original mask to the result:
\begin{equation}
S_{k_r}(x)=\left (\frac{G_{\sigma_1}*I}{G_{\sigma_1}*M}-\frac{G_{\sigma_2}*I}{G_{\sigma_2}*M}\right
) M
\label{eq:a9}
\end{equation}
The square of $S_{k_r}(x)$ is then integrated. The regions where the
image is not defined ($S_{k_r}(x)=0$), which do not contribute to the
variance,  are simply compensated for by counting all pixels
where $M(x)=0$ and making appropriate scaling:
\begin{equation}
V_{k_r,obs}=\frac{N}{N_{(M=1)}} \times \int S^2_{k_r}(x) d^nx ,
\label{eq:vobs}
\end{equation}
where $N=\int { d^nx}$ and $N_{(M=1)}=\int {M(x) d^nx}$. 

Comparing eq.~\ref{eq:vobs} and \ref{eq:expected} we get the final
estimate of the power density spectrum $\tilde{P}_{k_r}$ for a given
wave number $k_r$:
\begin{eqnarray}
\tilde{P}(k_r)=\frac{V_{k_r,obs}}{  \epsilon^2 n\left ( \frac{n}{2}+1
  \right) 2^{-\frac{n}{2}-1 } \pi^{\frac{n}{2}}
  k_r^n}=\frac{V_{k_r,obs}}{  \epsilon^2 \Upsilon(n) k_r^n},
\label{eq:rmsk}
\end{eqnarray}
where
\begin{eqnarray}
\Upsilon(n)=n\left ( \frac{n}{2}+1 \right) 2^{-\frac{n}{2}-1 } \pi^{\frac{n}{2}},
\end{eqnarray}
and 
$\Upsilon(n)=\frac{3}{4}\sqrt{\frac{\pi}{2}},~ \pi, \frac{15 \pi^{3/2}}{8
  \sqrt{2}}$ for $n=1,~2,~3$. 

The evaluation of the power density spectrum thus reduces to
eq.~\ref{eq:rmsk} for a set of values of $k_r$. 

\section{Normalization bias for a power law spectrum}
\label{ap:bias}
\begin{figure}
\includegraphics[width=0.45\columnwidth,bb=10 150 300 600]{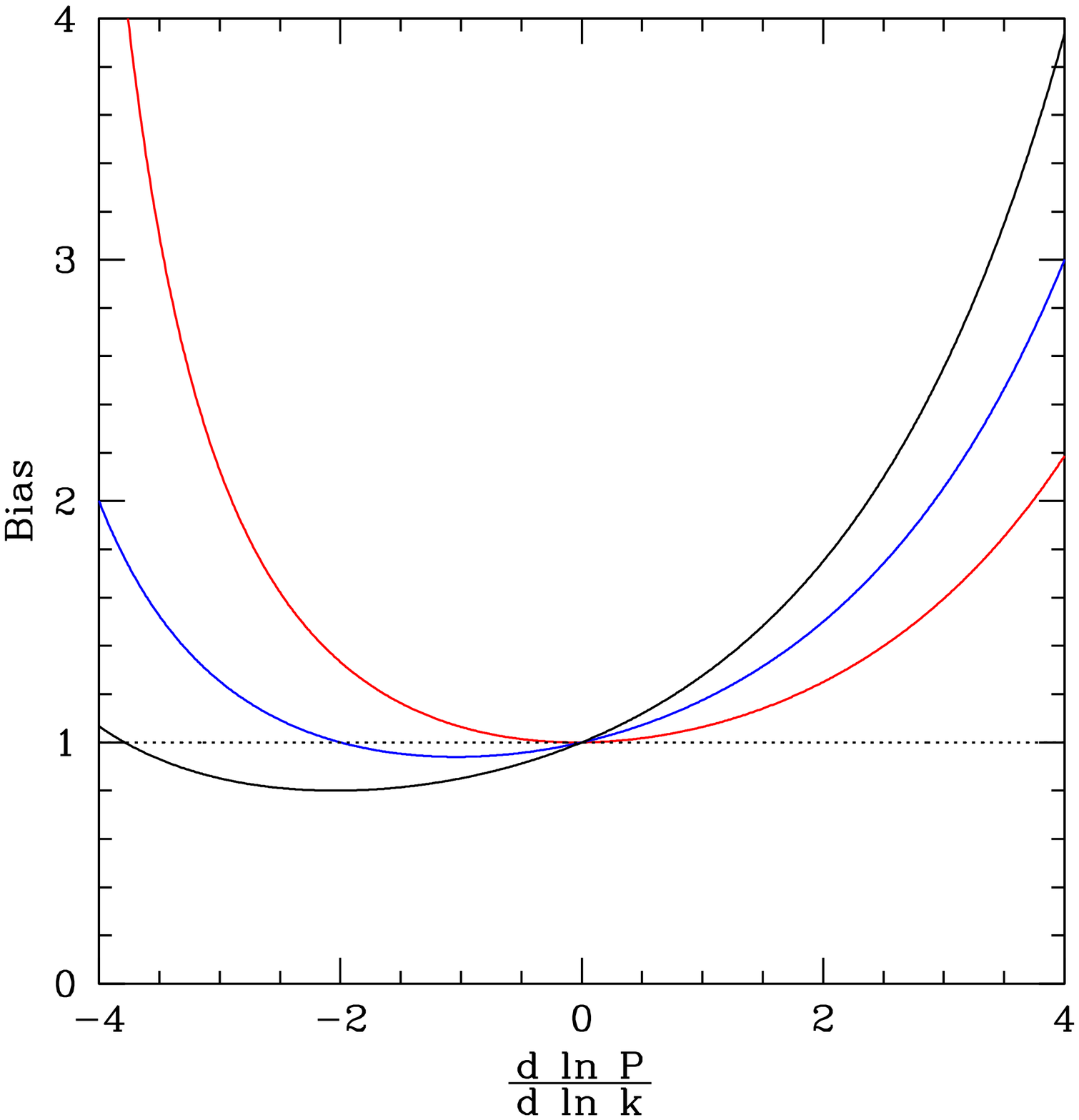}
\caption{ \label{fig:plb} Bias $\frac{\tilde{P}}{P}$ in the normalization of the recovered spectrum for a pure power law power spectrum, as a function of slope for different dimensions of the problem (red - 1D, blue - 2D, black - 3D)}. 
\end{figure}

For a pure power law spectrum $P(k)\propto k^{-\alpha}$ the expression
\ref{eq:van} can be easily evaluated and exact relation between
$\tilde{P}(k_r)$ and $P(k_r)$ can be written. The shape of the
spectrum is of course recovered correctly, while the normalization may
differ slightly, caused by moving $P(k)$ outside the integral in
eq.~\ref{eq:expected}:
\begin{equation}
\frac{\tilde{P}}{P}=2^{\alpha/2}\frac{\Gamma(\frac{n}{2}+2-\frac{\alpha}{2})}{\Gamma(\frac{n}{2}+2)}.
\end{equation}
Corresponding bias $\frac{\tilde{P}}{P}$ is shown in
Fig.~\ref{fig:plb}. One can see that for the most relevant problems (2D or
3D geometry, $\frac{d\ln P}{d\ln k}$ in the range [-3:0]) the bias is modest.
\label{ap:pl}

\label{lastpage}

\end{document}